\documentclass[12pt]{article}
\usepackage{lmodern}
\usepackage{multirow}
\usepackage{pgfplots}
\usepackage{tikz}
\usetikzlibrary{positioning}
\usepackage[export]{adjustbox}
\usetikzlibrary{arrows, shapes, snakes}
\usepackage{array}
\usepackage{rotating}
\usepackage{float}
\usepackage{amssymb}
\usepackage{amsmath}
\usepackage{tcolorbox}
\usepackage[round]{natbib}
\usepackage[colorlinks=true,linkcolor=black, citecolor=blue, urlcolor=blue]{hyperref}

\newcommand{\R}{\mathbb{R}}
\newcommand{\I}{\mathbf{1}}

\newtheorem{theorem}{Theorem}
\newtheorem{remark}{Remark}
\newtheorem{prop}{Proposition}
\newtheorem{algo-non}{Algorithm}
\DeclareMathOperator*{\argmin}{argmin} 

\definecolor{lightgray}{gray}{0.9}
\tikzstyle{startstop} = [ellipse, rounded corners, minimum width=3cm, minimum height=1cm,text centered, draw=blue!60, fill=blue!10]
\tikzstyle{arrow} = [thick,->,>=stealth]
\tikzstyle{rect} = [rectangle, rounded corners, minimum width=1.5cm, minimum height=1cm,text centered, draw=blue!60, fill=blue!10]

\begin{document}

\begin{center}
	\Large {{\scshape A clusterwise supervised learning procedure based on aggregation of distances}\bigskip\\ \large S. Has$^1$, A. Fischer$^1$ $\&$ M. Mougeot$^{1,2}$}
\end{center}
$^1$LPSM, Universit\'e de Paris, France\\
$^2$CMLA, Ecole Normale Sup\'erieure de Cachan $\&$ ENSIIE, France\medskip\\
\url{sothea.has@lpsm.paris}\\ \url{aurelie.fischer@lpsm.paris}\\ \url{mathilde.mougeot@ensiie.fr}

\begin{abstract}
Nowadays, many machine learning procedures are available on the shelve and may be used easily to calibrate predictive models on supervised data. However, when the input data consists of more than one unknown cluster, and when different underlying predictive models exist, fitting a model is a more challenging task. We propose, in this paper, a procedure in three steps to automatically solve this problem. The KFC procedure aggregates different models adaptively on data. The first step of the procedure aims at catching the clustering structure of the input data, which may be characterized by several statistical distributions. It provides several partitions, given the assumptions on the distributions. For each partition, the second step fits a specific predictive model based on the data in each cluster. The overall model is computed by a consensual aggregation of the models corresponding to the different partitions. A comparison of the performances on different simulated and real data assesses the excellent performance of our method in a large variety of prediction problems.
\end{abstract}

\noindent \emph{Keywords:}
Clustering, Bregman divergences, Aggregation, Classification, Regression, Kernel.

\noindent \emph{2010 Mathematics Subject Classification:} {62J99; 62P30; 68T05; 68U99}

\section{Introduction}
\label{sec:intro}
Machine learning tools and especially predictive models are today involved in a large variety of applications for the automated decision-making process such as face recognition, anomaly detection...
The final performance of a supervised learning model depends, of course, not only on the choice of the model but also on the quality of the dataset used to estimate the parameters of the model.
It is difficult to build an accurate model when some information is missing: the frequent expression ``garbage in, garbage out (GIGO)'' highlights that nonsense or incomplete input data produces nonsense output.

For some reasons, several fields useful for processing or understanding data may be missing. For instance, in hiring processes, the use of information about individuals, such as gender, ethnicity, place of residence, is not allowed for ethic reasons and to avoid discrimination. Similarly, when high school students apply for further studies in higher education, not every information can be considered for selection. Besides, the General Data Protection Regulation (GDPR) text regulates data processing in the European Union since May 2018. It strengthens the French Data Protection Act, establishing rules on the collection and use of data on French territory {\cite{tikkinen2018eu}}. As a result, contextual data that could characterize individuals a little too precisely is often missing in available databases. 
Moreover, in an industrial context, not all recorded fields are made available for data processing for confidentiality reasons. For example, in the automotive industry, GPS data could be a valuable tool to provide services such as predictive vehicle maintenance. However, it is difficult to use such data as they are extremely sensitive. 
To sum up, in various areas, databases containing individual information have to respect anonymization rules before being analyzed.

Mining such databases can then be a particularly complex task as some critical fields are missing.
In this context, the modalities of a missing qualitative variable correspond to several underlying groups of observations, which are a priori unknown but should be meaningful for designing a predictive model.
{In this case, the most common approach consists of using a two-step procedure: the clusters are computed in the first step and, in the second step, a predictive model is fit for each cluster.
This two-step procedure has already been used to approximate time evolution curves in the context of nuclear industry by \cite{auder2012projection},
to forecast electricity consumption using high-dimensional regression mixture models by \cite{devijver2015clustering},
or to cluster multi blocks before PLS regression by \cite{NBS15a}.
In a two-step procedure, the final performance of the model strongly depends on the first step.
Different configurations of clusters may bring various performances, and finding an appropriate configuration of clusters is not an easy task which often requires a deep data investigation and/or human expertise.}

To build accurate predictive models in situations where the contextual data are missing, and to eliminate an unfortunate choice of clusters, we propose, in this work, to aggregate several instances of the two-step procedures where each instance corresponds to a particular clustering. 
Our strategy is characterized by three steps, each is based on a quite simple procedure. 
The first step aims to cluster the input data into several groups and is based on the well-known $K$-means algorithm. As the underlying group structures are unknown and may be complex, a given Bregman divergence is used as a distortion measure in the $K$-means algorithm. 
In the second step, for each divergence, a very simple predictive model is fit per cluster. 
The final step provides an adaptive global predictive model by aggregating, thanks to a consensus idea introduced by \cite{mojirsheibani1999CombinedClass}, several models built for the different instances, corresponding to the different Bregman divergences (see also \cite{mojirsheibani2000KernelCombinedClass,balakrishnan2015SimpleMethod,cobra,mixcobra}). We name this procedure the {\it KFC} procedure for $K$-means/Fit/Consensus.

This paper is organized as follows. In Section \ref{section:not}, we recall some general definitions and notations about supervised learning. Section \ref{section:BregClust} is dedicated to Bregman divergences, their relationship with probability distributions of the exponential family, and $K$-means clustering with Bregman divergences. Section \ref{section:aggreg} presents the consensual aggregation methods considered, in classification and regression. The KFC procedure is detailed in Section \ref{section:model}. Finally, Sections \ref{section:simu} and \ref{section:appli} present several numerical results carried out on simulated and real data, showing the performance and the relevance of using our method. We also study the robustness of the procedure with respect to the number $K$ of clusters.

\section{Definitions and notations}\label{section:not}

We consider a general framework of supervised learning problems where the goal is to construct a predictive model using input data to predict the value of a variable of interest, also called response variable or output. 
Let $(X,Y)$ denote a random vector taking its values in $\R^d\times\mathcal{Y}$, where the output space $\mathcal{Y}$ is either $\{0,1\}$ (binary classification) or $\mathbb{R}$ (regression). Constructing a predictive model is finding a mapping $g:\mathcal{X}\rightarrow\mathcal{Y}$ such that the image $g(X)$ is ``close'' in some sense to the corresponding output $Y$. 
The space $(\R^d,\|\cdot\|)$ is equipped with the standard Euclidean metric. Let $\langle \cdot,\cdot\rangle$ denotes the associated standard inner product. Throughout, we take the convention $0/0=0$.

In classification problems, the performance of a predictor or classifier $g$ is usually measured using the misclassification error 
\begin{equation*}
\label{eq:1}
\mathcal{R}_C(g)=\mathbb{P}(g(X)\neq Y).
\end{equation*}
Similarly, the performance of a regression estimator $g$ is measured using the quadratic risk
\begin{equation*}
\label{eq:2}
\mathcal{R}_R(g)=\mathbb{E}\Big[\big(g(X)-Y\big)^2\Big].
\end{equation*}
In the sequel, $\mathcal R(g)$ describes the risk of a predictor $g$ without specifying the classification or regression case. A predictor $g^*$ is called optimal if
\begin{equation*}
\label{eq:3}
\mathcal{R}(g^*)=\inf_{g\in \mathcal{G}}\mathcal{R}(g)
\end{equation*}
where $\mathcal{G}$ is the class of all predictors $g:\R^d\rightarrow\mathcal{Y}$.
In regression, the optimal predictor is the regression function defined by $\eta(x)=\mathbb{P}(Y=1|X=x)$, whereas in binary classification the minimum is achieved by the Bayes classifier, given by $$g_B(x)=\begin{cases}1 &\mbox{ if } \eta(x)>1/2\\0 & \mbox{otherwise}.\end{cases}$$
Note that $\eta$ and, hence $g_B$, depend on the unknown distribution of $(X,Y)$.

In a statistical learning context, we observe independent and identically distributed random pairs $\{(X_1,Y_1),(X_2,Y_2),...,(X_n,Y_n)\}$ distributed as $(X,Y)$. The goal is to estimate the regression function $\eta$, or mimic the classifier $g_B$, based on the sample $D_n$.

We consider, in this work, situations where the input data $D_n$ may consist of several clusters and where there exist different 
underlying regression or classification models on these clusters.

\section{Bregman divergences and $K$-means clustering}
\label{section:BregClust}
Among all unsupervised learning methods, a well-known and widely used algorithm is the seminal $K$-means algorithm, based on the Euclidean distance,
see for example \cite{steinhaus1956division}, \cite{lloyd1982least}, \cite{linder2001LearningTheoreticMethod} or \cite{jain2010data}. This algorithm may be extended to other distortion measures, namely the class of Bregman divergences, \cite{benerjee2005ClusteringWithBD}. 

\subsection{Bregman Divergences}
\label{subsec:Breg}
Let $\phi:\mathcal{C}\rightarrow\mathbb{R}$ be a strictly convex and continuously differentiable function defined on a measurable convex subset $\mathcal{C}\subset\mathbb{R}^d$. Let $int(\mathcal{C})$ denote its relative interior. A Bregman divergence indexed by $\phi$ is a dissimilarity measure $d_{\phi}:\mathcal{C}\times int(\mathcal{C})\rightarrow\mathbb{R}$ defined for any pair $(x,y)\in \mathcal{C}\times int(\mathcal{C})$ by,
\begin{align}
\label{eq:1.10}
d_{\phi}(x,y)=\phi(x)-\phi(y)-\langle x-y,\nabla\phi(y)\rangle 
\end{align}
where $\nabla\phi(y)$ denotes the gradient of $\phi$ computed at a point $y\in int(\mathcal{C})$. A Bregman divergence is not necessarily a metric as it may not be symmetric and the triangular inequality might not be satisfied. However, it carries many interesting properties such as non-negativity, separability, convexity in the first argument, linearity in the indexed function, and the most important one is mean as minimizer property by \cite{banerjee2005optimality}.

\begin{prop}[\cite{banerjee2005optimality}]\label{prop:mean}
	Suppose $U$ is a random variable over an open subset $\mathcal{O}\subset\mathbb{R}^d$, then we have,
	$$\mathbb{E}[U]=\argmin_{x\in\mathcal{O}}\mathbb{E}[d_{\phi}(U,x)].$$
\end{prop}

In this article, we will consider four Bregman divergences, presented in \autoref{tab:1.1}: Squared Euclidean distance (Euclid), General Kullback-Leibler (GKL), Logistic (Logit) and Itakura-Saito (Ita) divergences. 

\begin{table}[h]
\scriptsize
\renewcommand{\arraystretch}{1.6}
\centering 
\begin{tabular}{| l | c | c | c |} 
\hline           
\textbf{BD} & \textbf{$\phi$} & \textbf{$d_{\phi}$} & \textbf{$\mathcal{C}$} \\ 
\hline 
Euclid & $\|x\|_2^2=\sum_{i=1}^d x_i^2$ & $\|x-y\|_2^2=\sum_{i=1}^d( x_i-y_i)^2$ & $\mathbb{R}^d$\\
\hline 
GKL & $\sum_{i=1}^d x_i\ln( x_i)$ & $\sum_{i=1}^d\Big[ x_i\ln(\frac{ x_i}{y_i})-( x_i-y_i)\Big]$ & $(0,+\infty)^d$\\
\hline 
Logit & $\sum_{i=1}^d\Big[ x_i\ln( x_i)+(1- x_i)\ln(1- x_i)\Big]$ &$\sum_{i=1}^d\Big[ x_i\ln(\frac{ x_i}{y_i})+(1- x_i)\ln(\frac{1- x_i}{1-y_i})\Big]$ & $(0,1)^d$\\
\hline 
Ita & $-\sum_{i=1}^d\ln( x_i)$ & $\sum_{i=1}^d\Big[\frac{ x_i}{y_i}-\ln(\frac{ x_i}{y_i})-1\Big]$ & $(0,+\infty)^d$\\
\hline             
\end{tabular}
\caption{Some examples of Bregman divergences.}%
\label{tab:1.1}
\end{table}

\subsection{Bregman Divergences and Exponential family}
\label{subsec:BregAndExp}
An exponential family is a class of probability distributions enclosing, for instance, Geometric, Poisson, Multinomial distributions, for the discrete case, Exponential, Gaussian, Gamma distribution, for the continuous case. More formally, an Exponential family $\mathcal{E}_{\psi}$ is a collection of probability distributions dominated by a $\sigma$-finite measure $\mu$ with density with respect to $\mu$ taking the following form:
\begin{align}
\label{eq:11}
f_{\theta}(x)=\exp(\langle\theta,T(x)\rangle-\psi(\theta)), \theta\in\Theta,
\end{align}
where $\Theta=\{\theta\in\mathbb{R}^d:\psi(\theta)<+\infty\}$ is the parameter space of natural parameter $\theta$, $T$ is called sufficient statistics and $\psi$ is called log-partition function.
The equation (\ref{eq:11}) is said to be {\it minimal} if the sufficient statistics $T$ is not redundant, that is,
if there does not exist any parameter $\alpha \ne 0$, such that $\langle \alpha,T(x)\rangle$ equals a constant, $\forall x\in\mathbb{R}^d.$ If the representation (\ref{eq:11}) is minimal and the parameter space $\Theta$ is open, then the family $\mathcal{E}_{\psi}$ is said to be {\it regular}. The relationship between a regular exponential family and Bregman divergence is given in the following theorem.
\begin{theorem}[\cite{benerjee2005ClusteringWithBD}] Each member of a regular exponential family corresponds to a unique regular Bregman divergence. If the distribution of a random variable $X$ is a member of a regular Exponential family $\mathcal{E}_{\psi}$ and if $\phi$ is the convex conjugate of $\psi$ defined by 
$$\phi(x)=\sup_{y}\{\langle x,y\rangle-\psi(y)\},$$
then there exists a unique Bregman divergence $d_{\phi}$ such that the following representation holds:
\begin{equation*}
\label{eq:12}
f_{\theta}(x)=\exp(\langle\theta,T(x)\rangle-\psi(\theta))=\exp(-d _{\phi}(T(x),\mathbb{E}[T(X)])+\phi(T(x))).
\end{equation*}\label{theo:bregexp}
\end{theorem}
Theorem \ref{theo:bregexp} and Proposition \ref{prop:mean} together provide a strong motivation for using $K$-means algorithm with Bregman divergences to cluster any sample distributed from the corresponding member of an exponential family.

We consider a set of $n$ input observations $\{ X_i\}_{i=1}^n$ distributed according to a law $f_{\theta}$, organized in $K$ clusters and $d_{\phi}$ is the associated Bregman divergence. Our goal is to find the centroids $\bf{c} = (c_1, \ldots, c_K)$ of the clusters minimizing the function
$$ W(f_{\theta}, {\bf c}) = \mathbb{E} \left[\min_{j =1,\ldots, K } d_{\phi} (X,c_j)  \right].$$
We propose the following $K$-mean clustering algorithm with the Bregman divergence $d_{\phi}$: 

\begin{algo-non}{\ }\\[-0.5cm]
\begin{enumerate}
\item\label{al:1} Randomly initialize the centroids 
$\{c_1^{(0)},c_2^{(0)},...,c_K^{(0)}\}$ among the data points.
\item\label{al:2} At iteration $r$: \textbf{For} $i=1,2,...,n,$ assign $ X_i^{(r)}$ to $k$-th cluster if 
$$d_{\phi}( X_i^{(r)},c_k^{(r)})=\min_{1\leq j\leq K} d_{\phi}( X_i^{(r)},c_j^{(r)})$$ 
\item\label{al:3} Denote by $C_k^{(r)}$ the set of points contained in the $k$-th cluster.\\
\textbf{For} $k=1,2,...,K,$ recomputes the new centroid by,
$$c_k^{(r+1)}=\frac{1}{|C_k^{(r)}|}\sum_{x\in C_k^{(r)}}x$$
\end{enumerate}
Repeat step \ref{al:2} and \ref{al:3} until a stopping criterion is met.
\end{algo-non}

In practice, it is well-known that the algorithm might get stuck at a local minimum if it begins with a bad initialization. A simple way to overcome this problem is to perform the algorithm several times with different initialization each time and to keep the partition minimizing the empirical distortion. In our version, in the event of ties, they are broken arbitrarily {and the associated empirical distortion is defined by
$$\widehat{W}({\bf c}) = \frac{1}{n}\sum_{i=1}^n\min_{1\leq k\leq K}d_{\phi}( X_i,c_k).$$}

For example:
\begin{itemize}
\item Poisson distribution with parameter $\lambda>0$: $X\sim{\cal P}(\lambda)$ has probability mass function: for any $k\in\{0,1,...\},\mathbb{P}(X=k)=e^{-\lambda}\frac{\lambda^k}{k!}$, corresponding to the 1-dimensional General Kullback-Leibler divergence defined by,
$$d_{\phi}(x,y)=x\ln\Big(\frac{x}{y}\Big)-(x-y),\forall x,y>0.$$
\item Exponential distribution with parameter $\lambda>0$: $X\sim{\cal E}(\lambda)$ has probability density function: for any $x>0,f_{\lambda}(x)=\lambda e^{-\lambda x}$, corresponding to the 1-dimensional Itakura-Saito divergence defined by,
$$d_{\phi}(x,y)=\frac{x}{y}-\ln\Big(\frac{x}{y}\Big)-1,\forall x,y>0.$$
\end{itemize}
See \cite{benerjee2005ClusteringWithBD} for more examples.

\section{Consensual aggregation methods}
\label{section:aggreg}

In this section, we describe the aggregation methods, based on a consensus notion, which will be used in the next section to build our global predictive model. The original combination idea was introduced by \cite{mojirsheibani1999CombinedClass} for classification (see also \cite{mojirsheibani2000KernelCombinedClass,mojirsheibani2006ComparisonStudy}) and adapted to the regression case by \cite{cobra}. We will also consider, in both classification and regression, a modified version of the consensual aggregation method introduced recently by \cite{mixcobra}.

\subsection{The original consensual aggregation}
\label{subsec:clasicAggre}
Several methods of combining estimates in regression and classification have been already introduced and studied.
\cite{leBlanc1996CombiningEstimationInRegAndClass} have proposed a procedure of combining estimates based on the linear combination of the estimated class of conditional probabilities, inspired on the "stacked regression" of \cite{breiman1996stacked}. 
Linear-type aggregation strategies, model selection and related problems have
been also studied by \cite{catoni2004statistical}, \cite{nemirovski2000topics}, \cite{yang2000combining}, \cite{yang2004aggregating},
and \cite{gyorfi2006distribution}. There are other related works by \cite{wolpert1992StackedGeneral}, and \cite{xu1992MethodOfCombiningClassAndApplication}.

In this paper, we will use a combining method introduced first in classification by \cite{mojirsheibani1999CombinedClass}, based on an idea of consensus. For a new query point $x\in\R^d$, the purpose is to search for data items $ X_i$, $i\in I$, such that all estimators to be combined predict the same label for $ X_i$ and $x$. The estimated label of $x$ is then obtained by a majority vote among the corresponding labels $Y_i$, $i\in I$. More formally, for $x\in\R^d$, let $\textbf{m}(x)=(m^{(1)}(x),...,m^{(M)}(x))$ denote the vector of the predictions for $x$ given by $M$ estimators. The combined estimator is defined by:
$$
	\label{eq:13}
	Comb_1^C(x)=
	\begin{cases} 1&\mbox{if }
	\displaystyle\sum_{i=1}^n
	\I_{\{\mathbf{m}( X_i)=\mathbf{m}(x)\}}
	\I_{\{Y_i=1\}}
	>\displaystyle\sum_{i=1}^n\I_{\{\mathbf{m}( X_i)=\mathbf{m}(x)\}}\I_{\{Y_i=0\}}\\
	0&\mbox{otherwise}.\end{cases}
$$

Under appropriate assumptions, the combined classifier is shown to asymptotically outperform the individual classifiers.
It is also possible to allow a few disagreements among the initial estimators. 

A regularized version, based on different kernels has been proposed in \cite{mojirsheibani2000KernelCombinedClass} (see also \cite{majid2016AsymptoticallyOptimal}). These smoother definitions are also a way not to require unanimity with respect to all the initial estimators, to lighten the effect of a possibly bad estimator in the list.

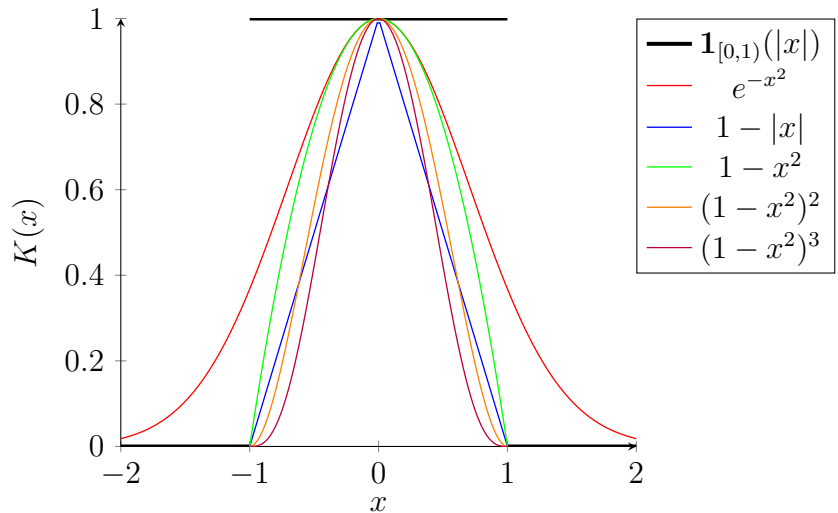
\begin{figure}[h]
	\centering
	\begin{tikzpicture}
	\begin{axis}[
	axis lines = left,
	xlabel = $x$,
	ylabel = {$K(x)$},
	legend style={at={(1,1)},anchor=north west},
	]
	
	\addplot [
	domain=-2:-1, 
	samples=100, 
	color=black,
	line width=1.5pt,
	]
	{0};
	\addplot [
	domain=1:2, 
	samples=100, 
	color=black,
	line width=1.5pt,
	]
	{0};
	
	\addplot [
	domain=-1:1, 
	samples=100, 
	color=black,
	line width=1.5pt,
	]
	{1};
	
	\addplot [
	domain=-2:2, 
	samples=100, 
	color=red,
	line width=0.5pt
	]
	{exp(-x^2)};
	\addplot [
	domain=-1:1, 
	samples=100, 
	color=blue,
	line width=0.5pt
	]
	{1-abs(x)};
	
	\addplot [
	domain=-1:1, 
	samples=100, 
	color=green,
	line width=0.5pt
	]
	{1-x^2};
	
	\addplot [
	domain=-1:1, 
	samples=100, 
	color=orange,
	line width=0.5pt
	]
	{(1-x^2)^2};
	
	\addplot [
	domain=-1:1, 
	samples=100, 
	color=purple,
	line width=0.5pt
	]
	{(1-x^2)^3};
	\legend{,,$\mathbf{1}_{[0,1)}(|x|)$,$e^{-x^2}$,$1-|x|$,$1-x^2$,$(1-x^2)^2$,$(1-x^2)^3$}
	\end{axis}
	\end{tikzpicture}
	\caption{The shapes of all kernels.}
	\label{fig:1}
\end{figure}

To simplify the notation, let $K$ be a positive decreasing kernel defined either on $\mathbb{R}_+$ or $\mathbb{R}^d$ to $\mathbb{R}_+$ then the kernel-based combined classifier is defined as follows:
$$
\label{eq:14}
Comb_2^C(x)=\begin{cases}1&\text{if }\displaystyle\sum_{i=1}^n(2Y_i-1)K_h\Big(d_{\mathcal{H}}(\mathbf{m}( X_i),\mathbf{m}(x))\Big)>0\\
0&\text{otherwise},
\end{cases}$$
where $d_{\mathcal{H}}$ stands for the Hamming distance (the number of disagreements between the components of $\mathbf{m}( X_i)$ and $\mathbf{m}(x)$), and $K_h(x)=K(x/h)$. We will consider the following kernels:
\begin{enumerate}
\item Gaussian kernel: for a given $\sigma>0$ and for all $x\in\mathbb{R}^d$, $$K(x)=e^{-\frac{\|x\|_2^{2}}{2\sigma^2}}.$$ 
\item Triangular kernel: for all $x\in\mathbb{R}^d$,
$$K(x)=(1-\|x\|_1)\mathbf{1}_{\{\|x\|_1\leq 1\}}.$$
where $\|.\|_1$ is the $\ell_1$-norm and is defined by: $\|x\|_1=\sum_{i=1}^d| X_i|$
\item Epanechnikov kernel: for all $x\in\mathbb{R}^d$,
$$K(x)=(1-\|x\|_2^2)\mathbf{1}_{\{\|x\|_2\leq 1\}}.$$
where $\|.\|_2$ is the $\ell_2$-norm and is defined by: $\|x\|_2=\Big(\sum_{i=1}^d X_i^2\Big)^{1/2}$
\item Bi-weight kernel: for all $x\in\mathbb{R}^d$,
$$K(x)=(1-\|x\|_2^2)^2\mathbf{1}_{\{\|x\|_2\leq 1\}}.$$
\item Tri-weight kernel: for all $x\in\mathbb{R}^d$,
$$K(x)=(1-\|x\|_2^2)^3\mathbf{1}_{\{\|x\|_2\leq 1\}}.$$
\end{enumerate}
These kernels are plotted in dimension 1 in Figure \ref{fig:1}, together with the uniform kernel corresponding to $Comb_1^C$.

In the regression case, mimicking the rule introduced in classification, the predictions will be required to be close to each other, in the sense of some threshold, with the predicted value obtained as a weighted average of the outputs of the selected data. The combined regression estimator, proposed in \cite{cobra} is given, for $x\in\R^d$, by
\begin{equation*}
\label{eq:16}
Comb^R_1(x)=\frac{1}{n}\sum_{i=1}^{n}W_{n,i}(x)Y_i
,\quad
W_{n,i}(x)=\frac{\prod_{\ell=1}^M\displaystyle\I_{\{|m^{(\ell)}( X_i)-m^{(\ell)}(x)|<\varepsilon\}}}{\sum_{j=1}^{n}\prod_{\ell=1}^M\I_{\{|m^{(\ell)}( X_j)-m^{(\ell)}(x)|<\varepsilon\}}}.
\end{equation*}
Once again, unanimity may be relaxed, for instance, if the distance condition is only required to be satisfied by a fraction $\alpha$ of the individual estimators:
$$W_{n,i}(x)=\frac{\displaystyle\I_{\big\{\sum_{\ell=1}^M\I_{\{|m^{(\ell)}( X_i)-m^{(\ell)}(x)|<\varepsilon\}}\geq M\alpha\big\}}}{\sum_{j=1}^{n}\I_{\big\{\sum_{\ell=1}^M\I_{\{|m^{(\ell)}( X_j)-m^{(\ell)}(x)|<\varepsilon\}}\geq M\alpha\big\}}}.$$
 It is shown that, when $\alpha\rightarrow 1$, the combined estimator asymptotically outperforms the different individual estimators. 
Here, we propose also to use a kernel version $Comb^R_2$, by setting:
\begin{equation*}
\label{eq:18}
W_{n,i}(x)=\frac{K_h(\mathbf{m}( X_i)-\mathbf{m}(x))}{\sum_{j=1}^{n}K_h(\mathbf{m}( X_j)-\mathbf{m}(x))}.
\end{equation*}

\subsection{Consensual aggregation combined to input distance}

An alternative definition of combined estimator suggests mixing the consensus idea with information about distances between inputs (\cite{mixcobra}). This is a way to limit the influence, if any, of a bad estimator; using at the same time information on the geometry of the inputs. In regression, the estimator is defined, for $x\in\R^d$, by
$$Comb_3^R(x)=\frac{1}{n}\sum_{i=1}^{n}W_{n,i}(x)Y_i
,\quad W_{n,i}(x)=\frac{K\Big(\frac{ X_i-x}{\alpha},\frac{\mathbf{m}( X_i)-\mathbf{m}(x)}{\beta}\Big)}{\sum_{j=1}^{n}K\Big(\frac{ X_j-x}{\alpha},\frac{\mathbf{m}( X_j)-\mathbf{m}(x)}{\beta}\Big)}.$$

In classification, by plug-in, we set 
\begin{equation*}
\label{eq:15}
Comb_3^C(x)=\begin{cases}1,&\text{if }\displaystyle\sum_{i=1}^{n}(2Y_i-1)K\Big(\frac{ X_i-X}{\alpha},\frac{\mathbf{m}( X_i)-\mathbf{m}(x)}{\beta}\Big)>0\\
0&\text{otherwise}.
\end{cases}
\end{equation*}

\section{The KFC procedure}\label{section:model}

We recall hereafter the three steps of the KFC strategy and specify the parameters chosen at each step.

\begin{enumerate}
\item{\it K-means.} The input data $X$ are first clustered using the $K$-means clustering algorithm with a chosen Bregman divergence. The choice of the number $K$ of clusters is discussed in the next Section \label{section:simu} where the numerical results on several examples are presented.
In this work, $M=4$ divergences are considered:
Squared Euclidean distance (Euclid), General Kullback-Leibler (GKL), Logistic (Logit) and Itakura-Saito (Ita) divergences, as
already defined in Section \ref{section:BregClust}.

\item{\it Fit.} For each Bregman divergence $m$ and for each cluster $k$, a dedicated predictive model, ${\cal M}_{m,k}$, is fit 
using the available observations, $1 \leq m \leq M$ and , $1 \leq k \leq K$.

In the numerical applications, we simply choose for regression models linear regression, whereas for the classification models, we choose logistic regression.
Much more complex models can be of course considered, but one of the main ideas of this paper, based on our modeling experience gained over several real-life projects, is that if the initial data are initially clustered <<in an appropriate way>> then the fit of the target variable can often be successfully computed with quite simple models in each group.

\item{\it Consensus.}  As neither the distribution nor the clustering structure of the input data is known, it is not clear in advance which divergence will be the most efficient. Thus, we propose to combine all the previous estimators, in order to take the best advantage of the clustering step. For the combination task, we use 
the different consensus-based procedures already described.
Practically, the different kernel bandwidths appearing in the combining methods are optimized on a grid, using cross-validation. 
\end{enumerate}

Once the candidate model, {which is the collection of all the local models constructed on the corresponding clusters}, is fitted, in order to make a prediction for a new observation $x$, we first affect $x$ to the closest cluster for each divergence, which yields one prediction per divergence, and then, performs the aggregation. The procedure is illustrated in Figure \ref{fig:2} below.

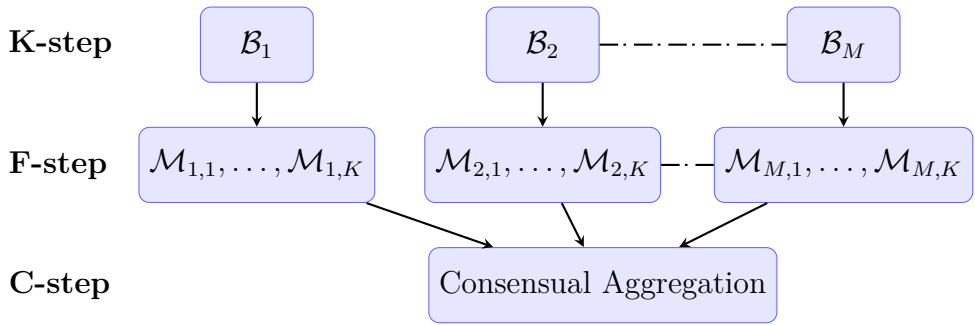
\begin{figure}[h!]
	\centering
	\begin{tikzpicture}[node distance=1.6cm]
	\node (bd1) [rect] {${\cal B}_1$};
	\node[left=of bd1, xshift=0.58cm] {\textbf{K-step}};
	\node (bd2) [rect, right of=bd1, xshift=2.2cm] {${\cal B}_2$};
	\node (bdm) [rect, right of=bd2, xshift=2.4cm] {${\cal B}_M$};
	\node (cp1) [rect, below of=bd1] {${\cal M}_{1,1},\dots,{\cal M}_{1,K}$};
	\node [left=of cp1, xshift=1.33cm] {\textbf{F-step}};
	\node (cp2) [rect, below of=bd2] {${\cal M}_{2,1},\dots,{\cal M}_{2,K}$};
	\node (cpm) [rect, below of=bdm] {${\cal M}_{M,1},\dots,{\cal M}_{M,K}$};
	\node (com)[rect, below of=cp2, xshift=0.8cm] {Consensual Aggregation};
	\node[left=of com, xshift=-2.48cm] {\textbf{C-step}};
	\draw [thick,dash pattern={on 7pt off 2pt on 1pt off 3pt}] (bd2) -- (bdm);
	\draw [thick,dash pattern={on 7pt off 2pt on 1pt off 3pt}] (cp2) -- (cpm);
	\draw [arrow] (bd1) -- (cp1);
	\draw [arrow] (bd2) -- (cp2);
	\draw [arrow] (bdm) -- (cpm);
	\draw [arrow] (cp1) -- (com);
	\draw [arrow] (cp2) -- (com);
	\draw [arrow] (cpm) -- (com);
	\end{tikzpicture}
	\caption{The main steps of the model construction: for each Bregman divergence ${\cal B}_m$, one model ${\cal M}_{m,k}$ is fit per cluster $k$, then the models corresponding to the different divergences are combined.}
	\label{fig:2}
\end{figure}


\section{Simulated data}
\label{section:simu}
In this section, we analyze the behavior of the strategy on several simulated datasets in both classification and regression problems.

\subsection{Description}
In both cases of classification and regression problems, we simulate 5 different kinds of datasets.
We consider 2-dimensional datasets where the two predictors $(X_1, X_2)$ are simulated according to Exponential, Poisson, Geometric and Gaussian distribution respectively. The remaining dataset is 3-dimensional, with predictors $(X_1, X_2, X_3)$, distributed according to Gaussian distribution. 
Each simulated training and testing dataset contains respectively $1500$ and $450$ data points. 
Each dataset consists of $K=3$ balanced clusters; each cluster contains $500$ observations for training and $150$ for testing. {Note that this choice of $K=3$ clusters is to illustrate the procedure and performance of our algorithm. Various complementary studies with different number of clusters showed that similar results held.}

The different distribution parameters used in the simulations are listed in \autoref{tab:2}. {Each cell of the table contains the parameters of each distribution at the corresponding cluster for the input variables $(X_1,X_2)$ or $(X_1,X_2,X_3)$.}

\begin{table}[h!]
\centering 
\begin{tabular}{l l l l l} 
\hline\hline            
\textbf{Distribution} & \textbf{Cluster 1} & \textbf{Cluster 2} & \textbf{Cluster 3}\\[0.5ex] 
\hline 
Exponential: $\lambda$ & $0.05;0.5$ & $0.5;0.05$ & $0.1;0.1$\\[1ex]
Poisson: $\lambda$  & $3;11$ & $10;2$ & $13;12$\\[1ex]
Geometric: $p$ & $0.07;0.35$ & $0.55;0.07$ & $0.15;0.15$\\[1ex]
2D Normal: $\begin{cases}\mu\\ \sigma\end{cases}$ & $\begin{cases}4;12\\ 1;1\end{cases}$ & $\begin{cases}22;9\\ 2;1\end{cases}$ & $\begin{cases}10;5\\ 2;2\end{cases}$\\[3ex]
{3D Normal} $\begin{cases}\mu\\ \sigma\end{cases}$ & $\begin{cases}6;14;6\\ 1;2;1\end{cases}$ & $\begin{cases}5;10;15\\2;1;2\end{cases}$ & $\begin{cases}8;6;14\\1;1;2\end{cases}$\\[1ex]
\hline             
\end{tabular}
\caption{Parameters of the simulated data.}
\label{tab:2}
\end{table}

For the regression cases,  the target observation $Y_i$ belonging to cluster $k$, is computed by $Y_i^k = \beta_0^k + \sum \beta_j^k  X_i ^k+ \epsilon_i$
where $X_{i}^k=(X_{i,j}^k)_{j=1,...,d}$ is the input observation of dimension $d$, 
$\beta^k=(\beta^k_{j})_{j=1,...,d}$ the parameters of cluster $k$, $1 \leq k \leq K$, $d=2$ or $d=3$ 
and $\epsilon_i \sim \mathcal{N}(0,10)$. 

For classification cases, the target observation belonging to cluster $k$, is computed by 
$Y_i^k = 0$ if $\frac{1-e^{\beta_0^k + \sum \beta_j^k  X_i ^k + \epsilon_i}}{1+ e^{\beta_0^k + \sum \beta_j^k  X_i ^k + \epsilon_i}} \leq 0$
and $\epsilon_i \sim \mathcal{N}(0,10)$. 

\begin{table}[h!]
\centering 
\begin{tabular}{lllllll} 
\hline\hline            
 & \textbf{Cluster 1} & \textbf{Cluster 2} & \textbf{Cluster 3}\\[0.5ex] 
 & ($k=1$) & ($k=2$) & ($k=3$) \\
\hline 
{2D} $(\beta_1^k,\beta_2^k)$ & $(-8,3)$ & $(-6,-5)$ & $(5,-7)$\\
\hline                  
{3D} $(\beta_1^k,\beta_2^k,\beta_3^k)$ & $(-10,3,7)$ & $(7,5,-12)$ & $(6,-11,10)$\\
\hline         
\end{tabular}
\caption{The coefficients of the simulated models.}
\label{tab:3}
\end{table}

\noindent In regression problems, we choose the intercepts $(\beta_0^1,\beta_0^2,\beta_0^3)=(-15,25,-10)$ for the 3 clusters. 
For classification, we study cases where each cluster has the same number of observations form the 2 labels.
In order to balance the positive and negative points in classification cases, we choose intercepts so that the hyperplane defined by the input data within each cluster is centered at zero. Therefore, after applying the sigmoid transformation, we would have a balance between the two classes within each cluster. This can be done as follows.
\begin{itemize}
\item Compute $\alpha_j^k$: the conditional average of the $j$-th input variable falling into the $k$-th cluster which is defined by
$$\alpha_j^k=\frac{1}{|C_j^k|}\sum_{x\in C_j^k} x$$
where $C_j^k\subset  X_j$ is the subset of the $j$-th input variable that are contained in the $k$-th cluster.
\item The intercept of the $k$-th cluster for $k\in\{1,2,3\}$ is given by,
$$
\beta_0^k=-\langle \beta^k,\alpha^k\rangle=\sum_{j=1}^d\alpha_j^k\beta_j^k,\ \text{for } d=2\text{ or }d=3
$$
\end{itemize} 

\begin{figure}[!h]
\centering
\includegraphics[width=10cm]{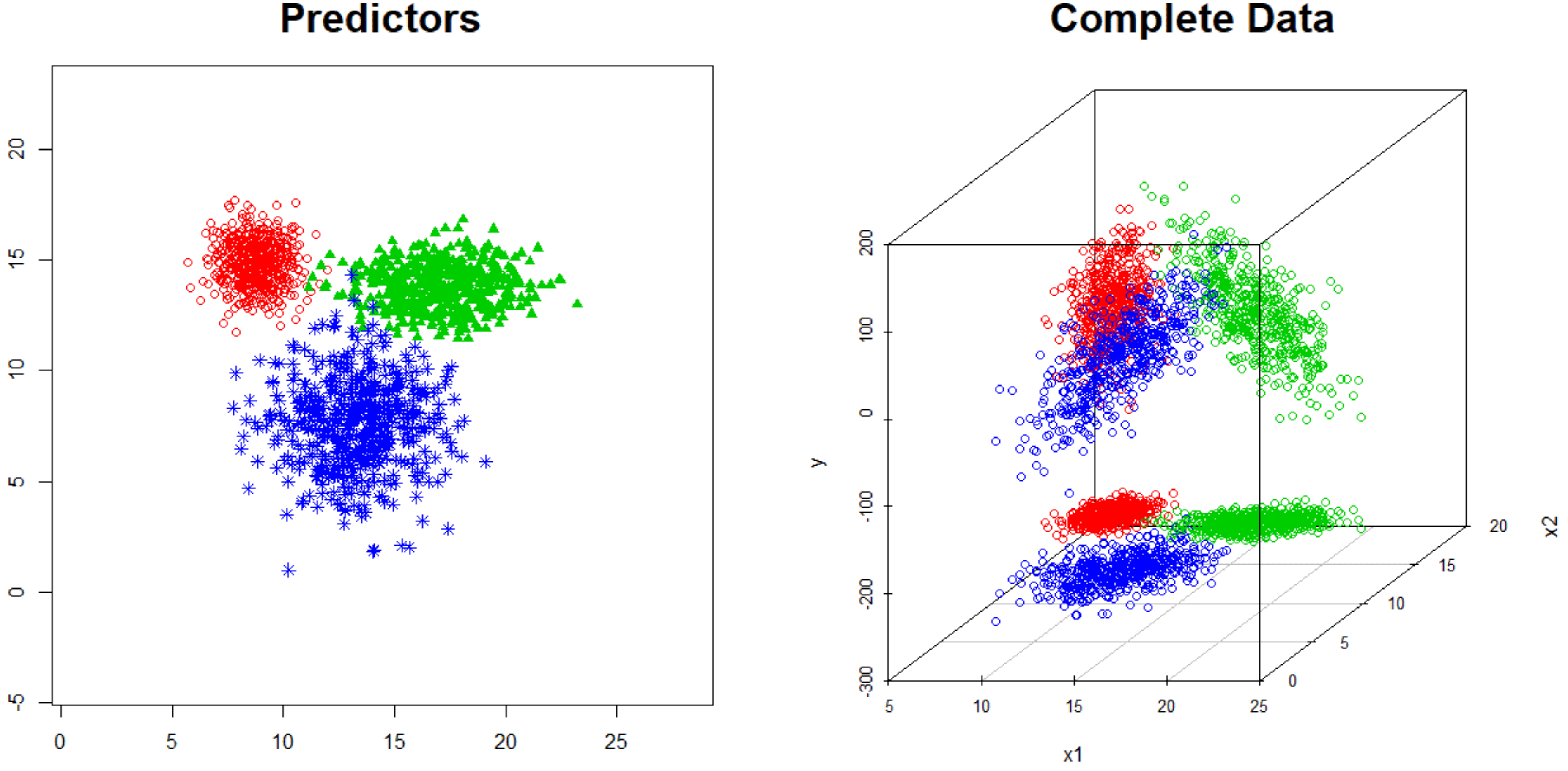}
\caption{An example of simulated data in regression problem with Gaussian predictors.}
\label{fig:3}
\end{figure}

\begin{remark}
Note that in our simulations, the simulated samples might fall outside the domain $\mathcal{C}$ for some Bregman divergences for instance, the logistic one which can handle only data points in $(0,1)^d$. In practice, we can solve this problem by normalizing our original samples using the $\ell_1$-norm $\|.\|_1$, i.e., $ X_i\rightarrow\tilde{X}_i= X_i/\| X_i\|_1$. Moreover, we ignored those negative data points or added a suitable constant in order to avoid negativity.
\end{remark}

Each performance is computed over 20 replications of the corresponding dataset.

\subsection{Normalized Mutual Information}
\label{subsec:clasicAggre}
Before analyzing the performances of our combined estimators, it is interesting to take a look at the performances of the clustering algorithm itself with different Bregman divergences. Even though this is not possible in practice, the clustering structure is here available in our simulations. We use a correlation coefficient between partitions proposed by \cite{strehl2002ClusterEnsembles} known as Normalized Mutual Information (NMI). Let $S=\{S_j\}_{j=1}^K$ and $S'=\{S_{\ell}'\}_{\ell=1}^K$ be two partitions of $n$-point observations. Let $n_j$, $n_{\ell}'$ and $n_{j,\ell}$ denote the number of observations in $S_j\in S, S_{\ell}'\in S'$ and $S_j\cap S_{\ell}'$ respectively. Then, the NMI of the two partitions $S$ and $S'$ is given by

\begin{equation*}
\label{eq:20}
\rho(S,S')=\frac{\sum_{j=1}^K\sum_{\ell=1}^Kn_{j,\ell}\log\Big(\frac{n.n_{j,\ell}}{n_jn_{\ell}'}\Big)}{\sqrt{\Big(\sum_{j=1}^Kn_j\log\Big(\frac{n_j}{n}\Big)\Big)\Big(\sum_{\ell=1}^Kn_{\ell}'\log\Big(\frac{n_{\ell}'}{n}\Big)\Big)}}.
\end{equation*}

\noindent This criterion allows us to compare the observed partition given by the clustering algorithm to the expected (true) one. We have $0\leq \rho(S,S')\leq 1$ for any partitions $S$ and $S'$. The closer coefficient to $1$, the better the result of the clustering algorithm. 

\begin{table}[H]
	\centering 
	\begin{tabular}{l | c c c c} 
		\hline\hline            
		Distributions & Euclidean & GKL & Logistic & Itakura-Saito
		\\ [0.5ex]  
		\hline 
		\multirow{2}{*}{Exponential} & $17.77$ & $24.79$ & $60.42$ & $\textcolor{blue}{\textbf{76.61}}$ \\
		& $(1.53)$ & $(2.26)$ & $(1.35)$ & $(1.82)$ \\ \cline{1-5}
		\multirow{2}{*}{Poisson} & $88.26$ & $\textcolor{blue}{\textbf{92.24}}$ & $68.19$ & $83.53$ \\
		& $(1.16)$ & $(1.41)$ & $(1.47)$ & $(9.85)$\\ \cline{1-5}
		\multirow{2}{*}{Geometric} & $53.6	1$ & $86.06$ & $\textcolor{blue}{\textbf{87.31}}$ & $81.16$\\
		& $(1.86)$ & $(10.04)$ & $(0.82)$ & $(1.56)$ \\ \cline{1-5}
		\multirow{2}{*}{2D Normal} & $\textcolor{blue}{\textbf{97.89}}$ & $97.46$ & $69.56$ & $94.81$ \\
		& $(0.89)$ & $(0.99)$ & $(1.41)$ & $(1.29)$\\ \cline{1-5}
		\multirow{2}{*}{3D Normal} & $\textcolor{blue}{\textbf{91.55}}$ & $91.19$ & $89.22$ & $89.95$\\
		& $(1.31)$ & $(1.22)$ & $(1.57)$ & $(1.66)$\\
		\hline             
	\end{tabular}
	\caption{Average Normalized Mutual Information ($1$ unit $=10^{-2}$).}%
	\label{tab:4}
\end{table}

\begin{figure}[H]
\centering
\includegraphics[width=13.5cm, height=7.15cm]{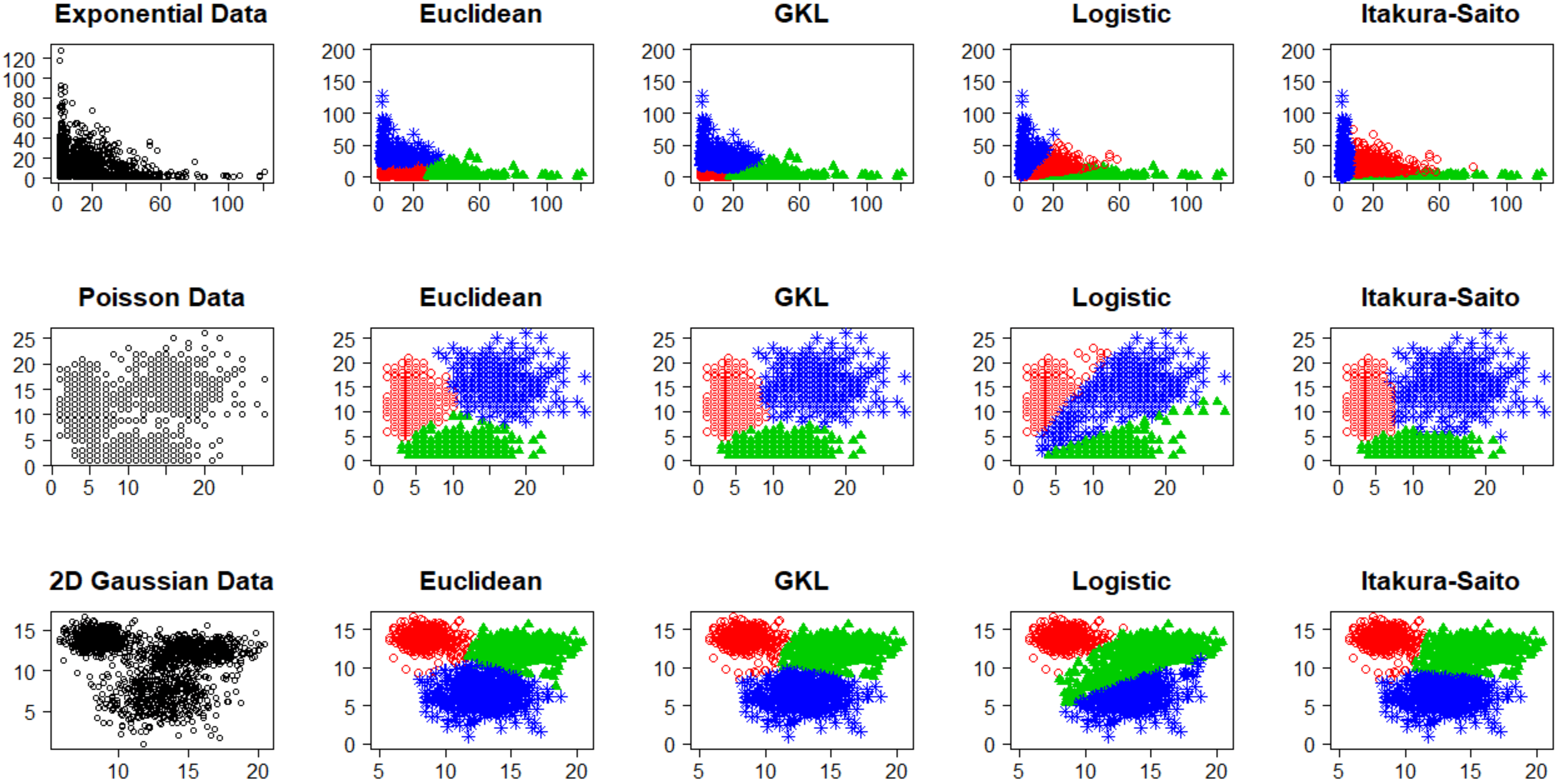}
\caption{Partitions obtained via $K$-means with Bregman divergences.}\label{fig:4}
\end{figure}

\autoref{tab:4} above contains the average NMI { over $20$ runs of $K$-means clustering algorithm performed on each simulated dataset.} The associated standard deviations are provided in brackets. The out-performance of each case is highlighted in blue. {Note that the results in the Table \ref{tab:4} recover the expected relation between distributions and Bregman divergences as discussed in Section~\ref{subsec:BregAndExp}}.
\autoref{fig:4} illustrates the computed partitions for one run simulation using $K$-means algorithm with Bregman divergences.

\subsection{Numerical results}
\label{subsec:numeric}
This section analyzes the ability of the KFC procedure for classification or regression on the five simulated examples described in section \ref{section:simu}.
Each example is simulated 20 times. For each run, the error obtained using the KFC procedure is computed on the Test dataset;
the classification error is evaluated using the misclassification rate and the regression error the Root Mean Square Error.
The average and the standard deviation (in bracket) of the errors computed over the 20 runs are provided in the result tables.
In order to compare the benefit of the consensual aggregation of KFC procedure, we evaluate the performance of the model
on the test data in different situations.
First, without any preliminary clustering (i.e. considering only one cluster), the corresponding errors are reported in the column block named "Single" in the different graphs or tables. Second, considering a preliminary clustering using one given divergence.
In this case, the corresponding errors are reported in the column block named "Bregman divergence" in different tables.
The four columns named Euclid, GKL, Logistic and Ita contain the results of the 4 individual estimators corresponding to the 4 chosen Bregman divergences. Last, the errors computed with the KFC procedure are presented with several kernels in the block named ``Kernel'' which consists of six columns named Unif, Epan, Gaus, Triang, Bi-wgt and Tri-wgt standing for Uniform, Epanechnikov, Gaussian, Triangular, Bi-weight and Tri-weight kernel (procedures $Comb_1, Comb_2$).
The KFC procedure is also evaluated taking into account the inputs ($Comb_3$), and the corresponding results are provided in the second row of each distribution.

For each table, the first column of each row mentions the names of the simulated datasets where Exp, Pois, Geom, 2D Gauss, and 3D Gauss stand for Exponential, Poisson, Geometric, 2-dimensional and 3-dimensional Gaussian datasets respectively.

For each distribution, we highlight the out-performance of the individual estimators in bold font and the two kinds of combining methods in boldfaced blue $Comb_1, Comb_2$) and red ($Comb_3$) respectively. In each simulation, we consider $300$ values of smoothing parameter $h$ or $\varepsilon$ on the grid $\{10^{-300},...,5\}$ for $Comb_1$ and $Comb_2$, and consider $50\times50$ values of parameters $(\alpha,\beta)\in\{10^{-300},...,10\}^2$ for $Comb_3$.

\subsubsection{Classification}
\label{subsec:classi}

\begin{figure}[ph!]
\centering
\includegraphics[width = 6.5cm, height = 4.5cm]{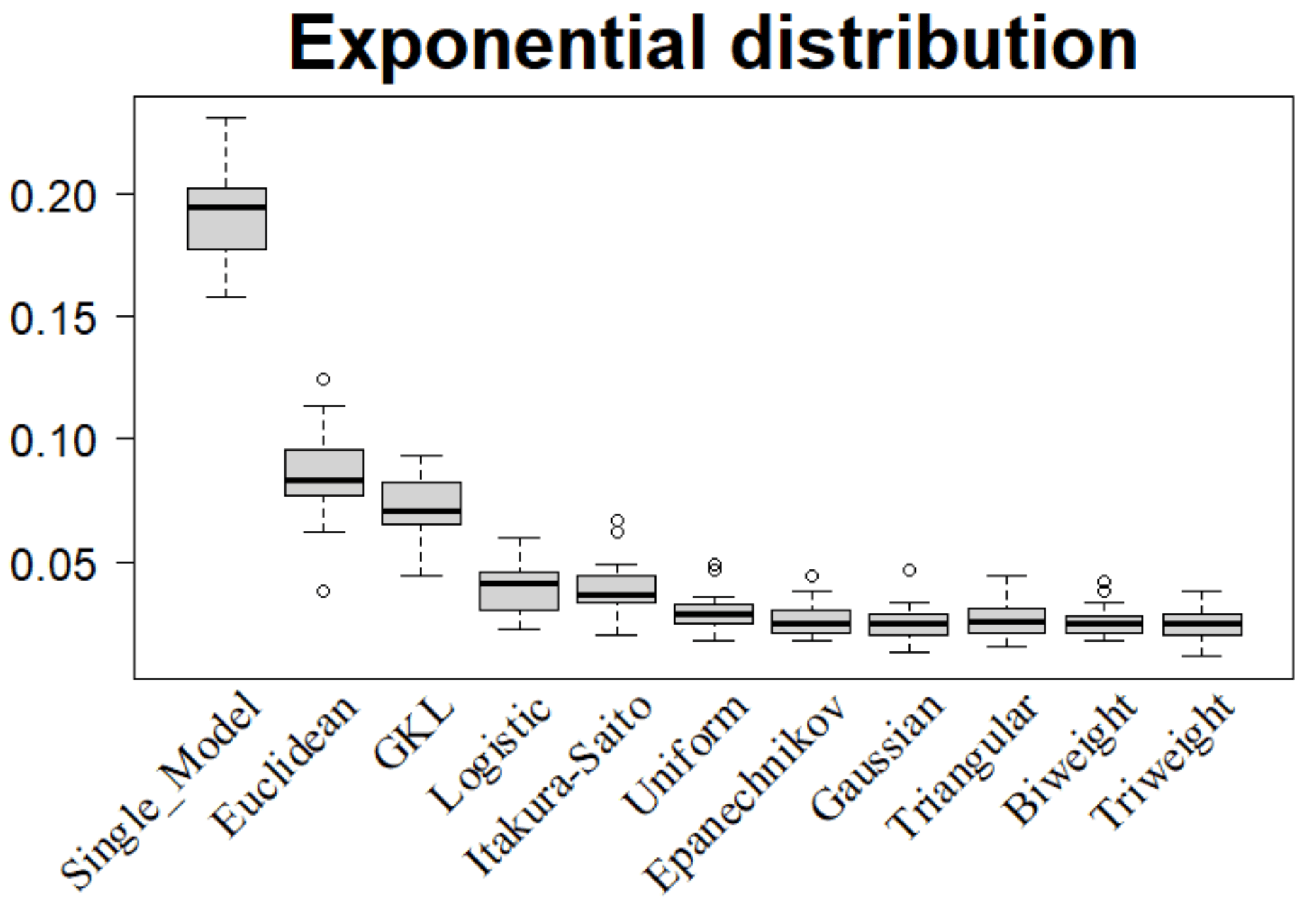}
\includegraphics[width = 6.5cm, height = 4.5cm]{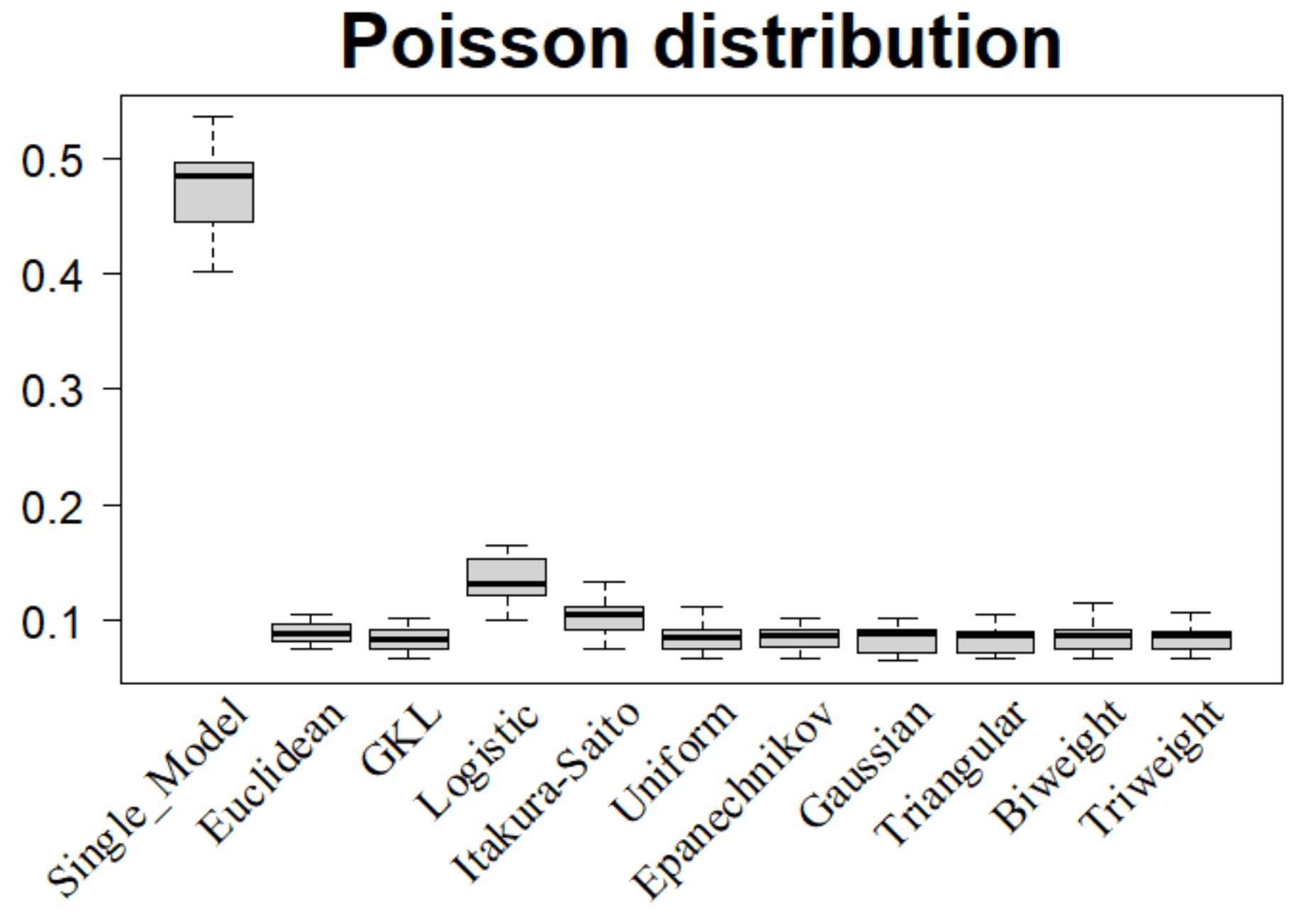}
\includegraphics[width = 6.5cm, height = 4.5cm]{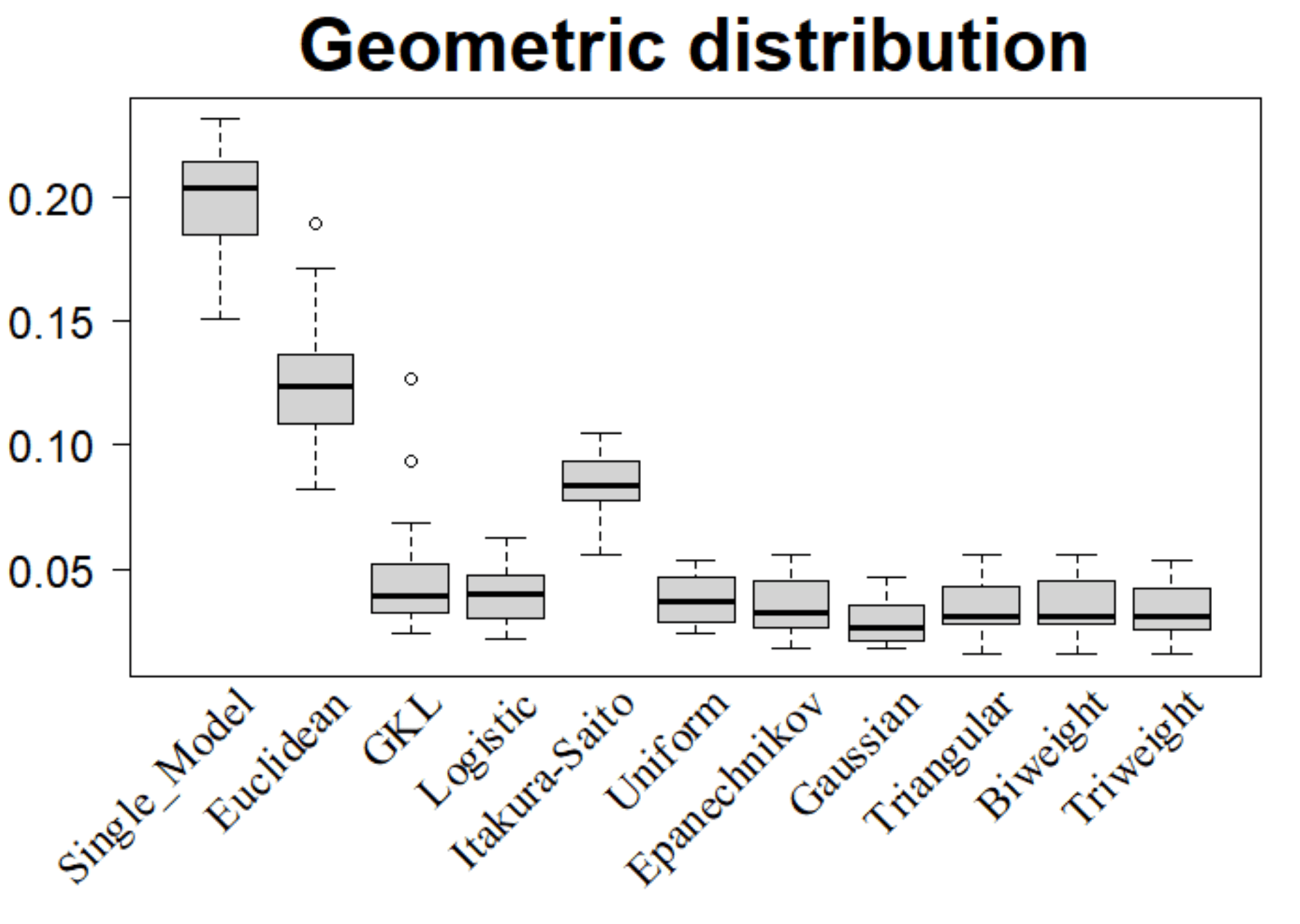}
\includegraphics[width = 6.5cm, height = 4.5cm]{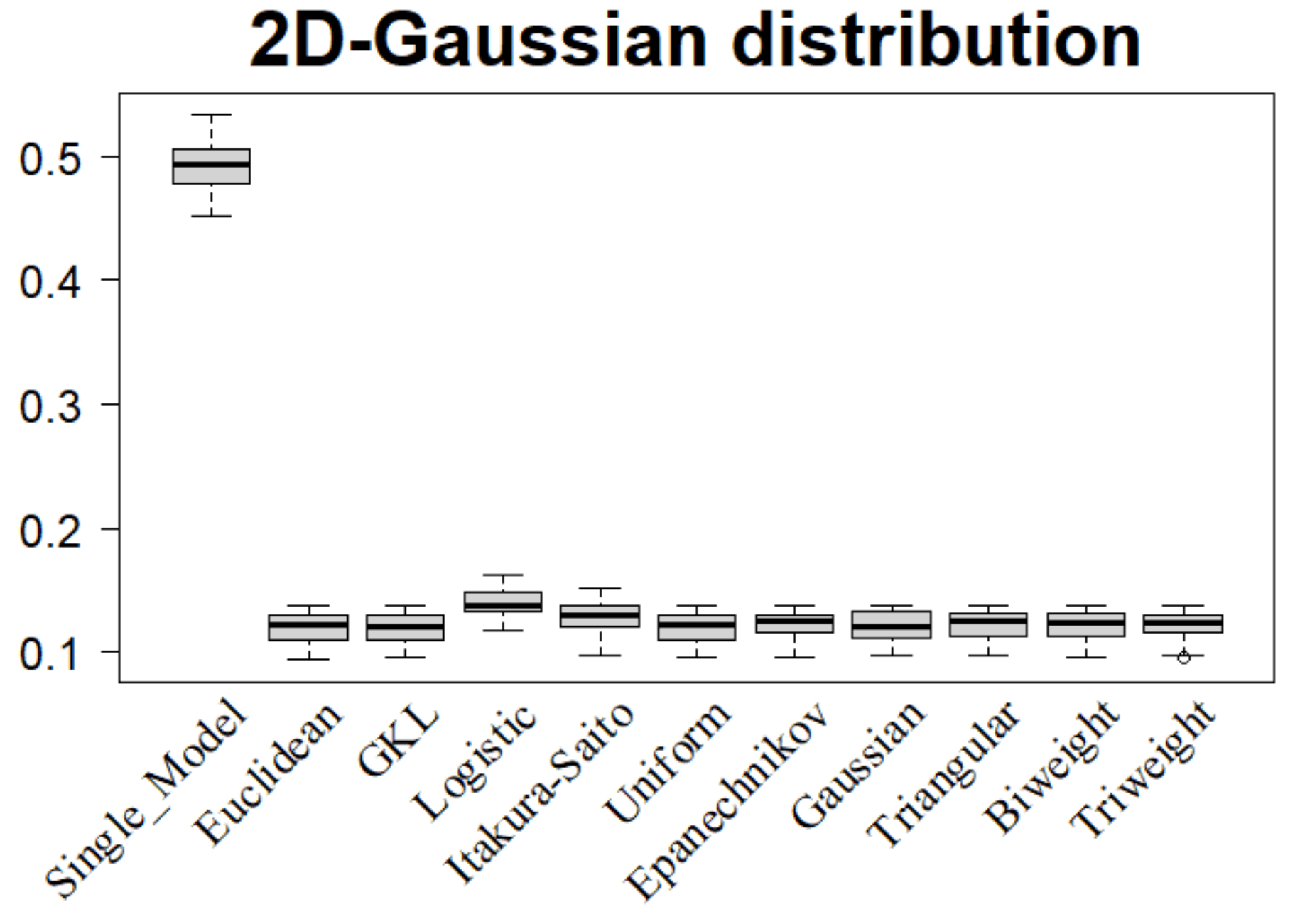}
\includegraphics[width = 6.5cm, height = 4.5cm]{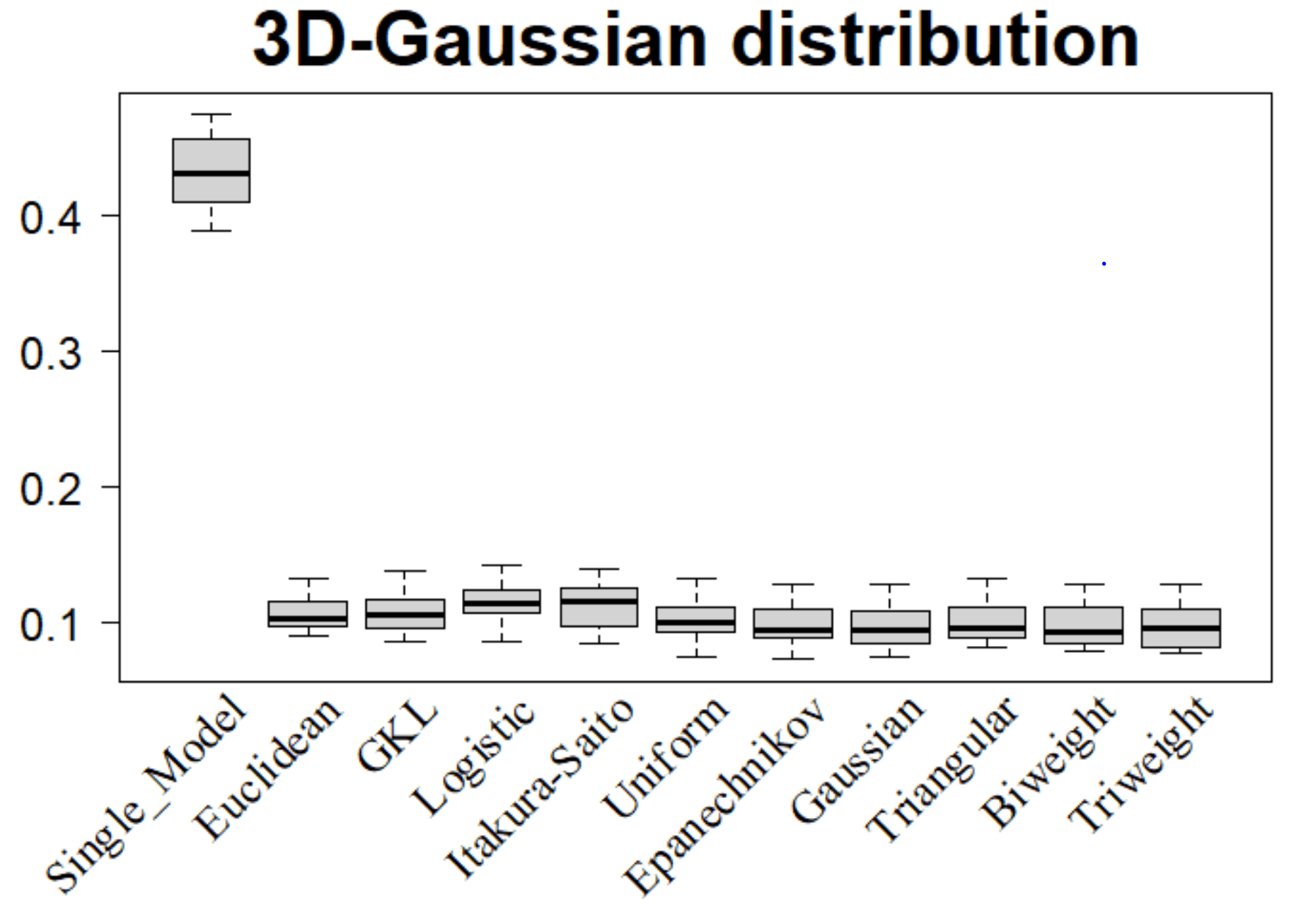}
\caption{Boxplots of misclassification error of $Comb_3^C$.}
\label{fig:6}
\end{figure}

\autoref{tab:5} below contains the results of misclassification errors computed on the different kinds of simulated datasets. We observe that the results of all individual estimators in the second block seem to agree with the results of NMI provided in \autoref{tab:4}. Of course, all models built after a clustering step outperform the simple model of the first block. 
\begin{figure}[ph!]
\centering
\includegraphics[width = 6.5cm, height = 4.6cm]{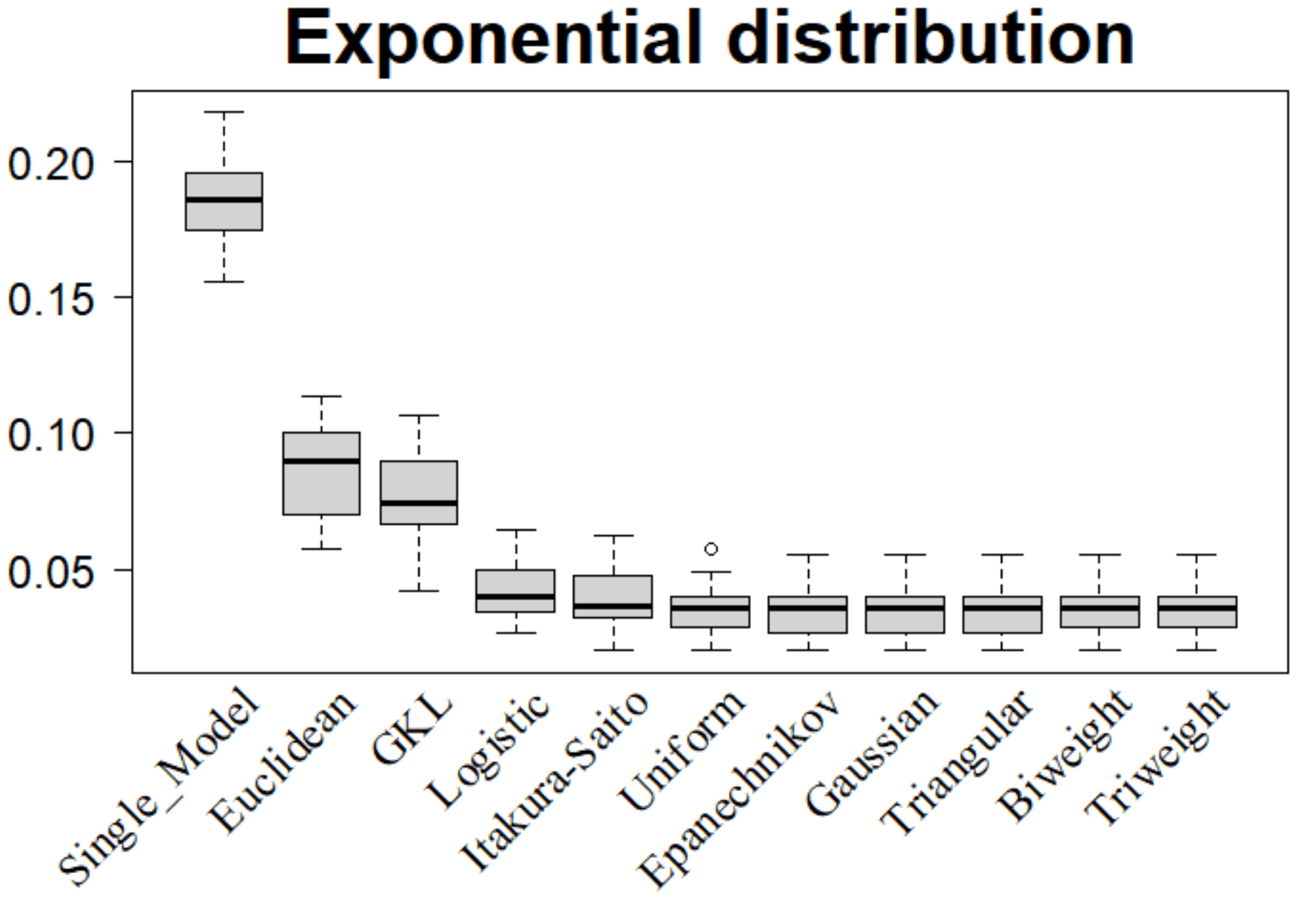}
\includegraphics[width = 6.5cm, height = 4.6cm]{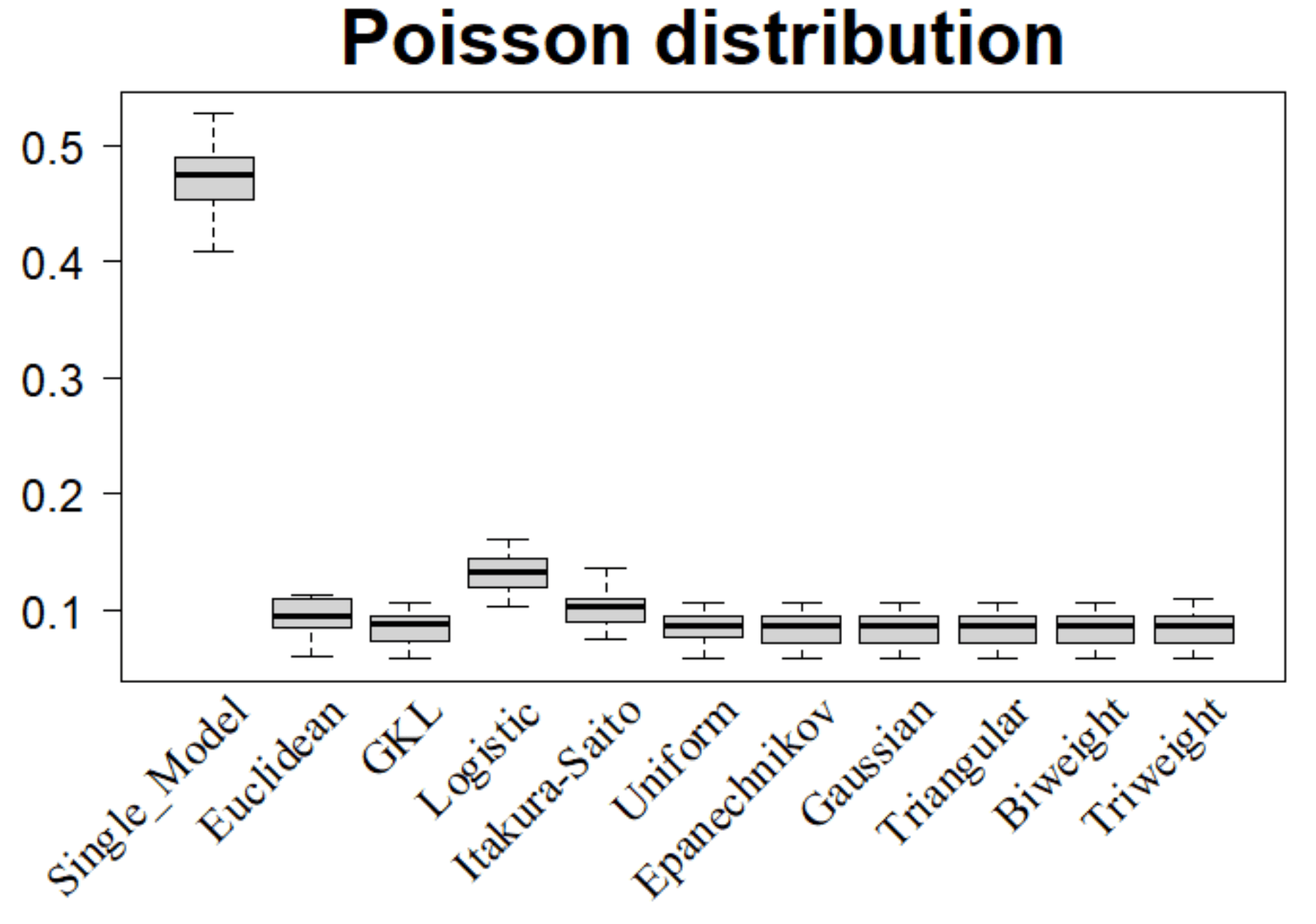}
\includegraphics[width = 6.5cm, height = 4.6cm]{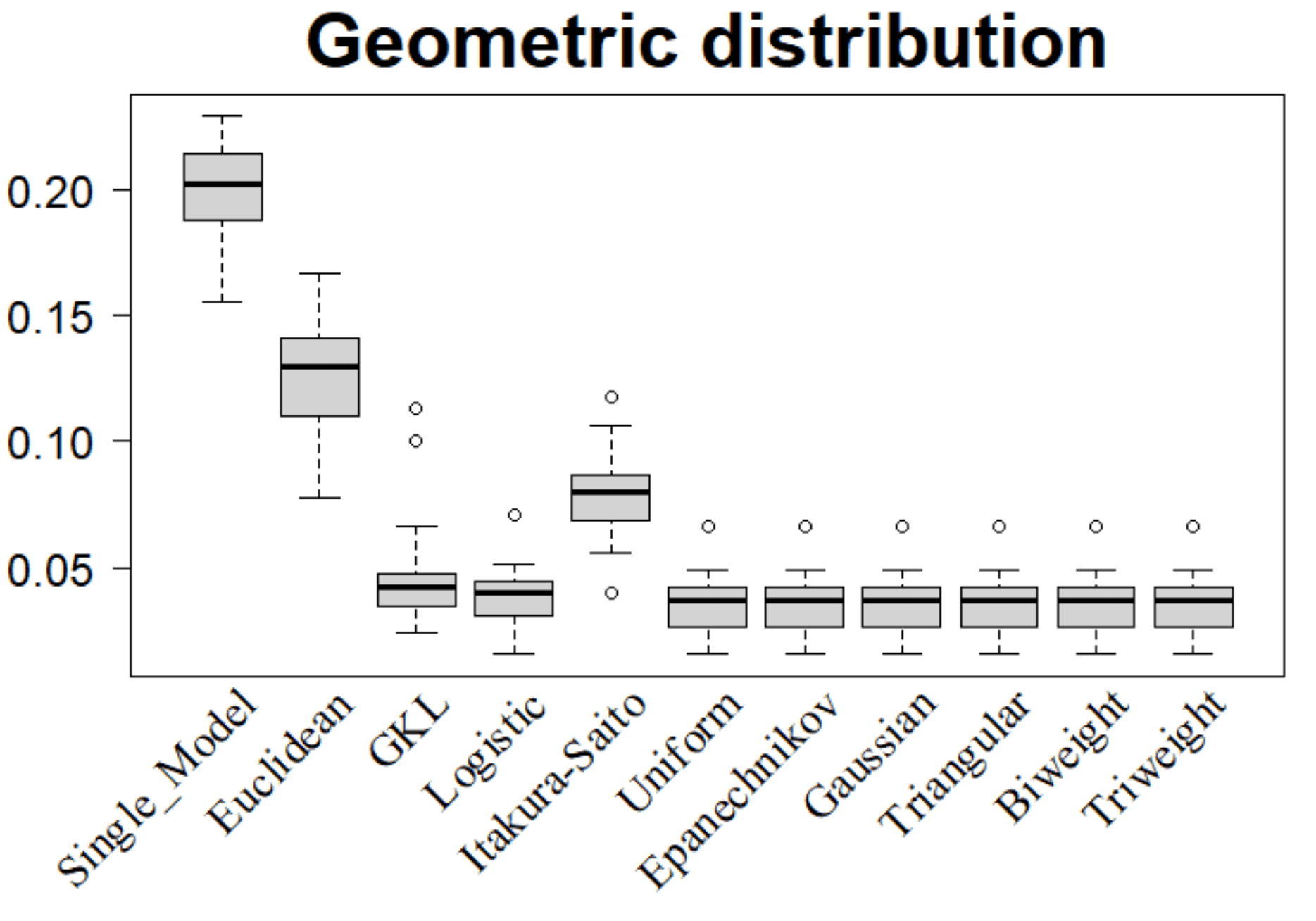}
\includegraphics[width = 6.5cm, height = 4.6cm]{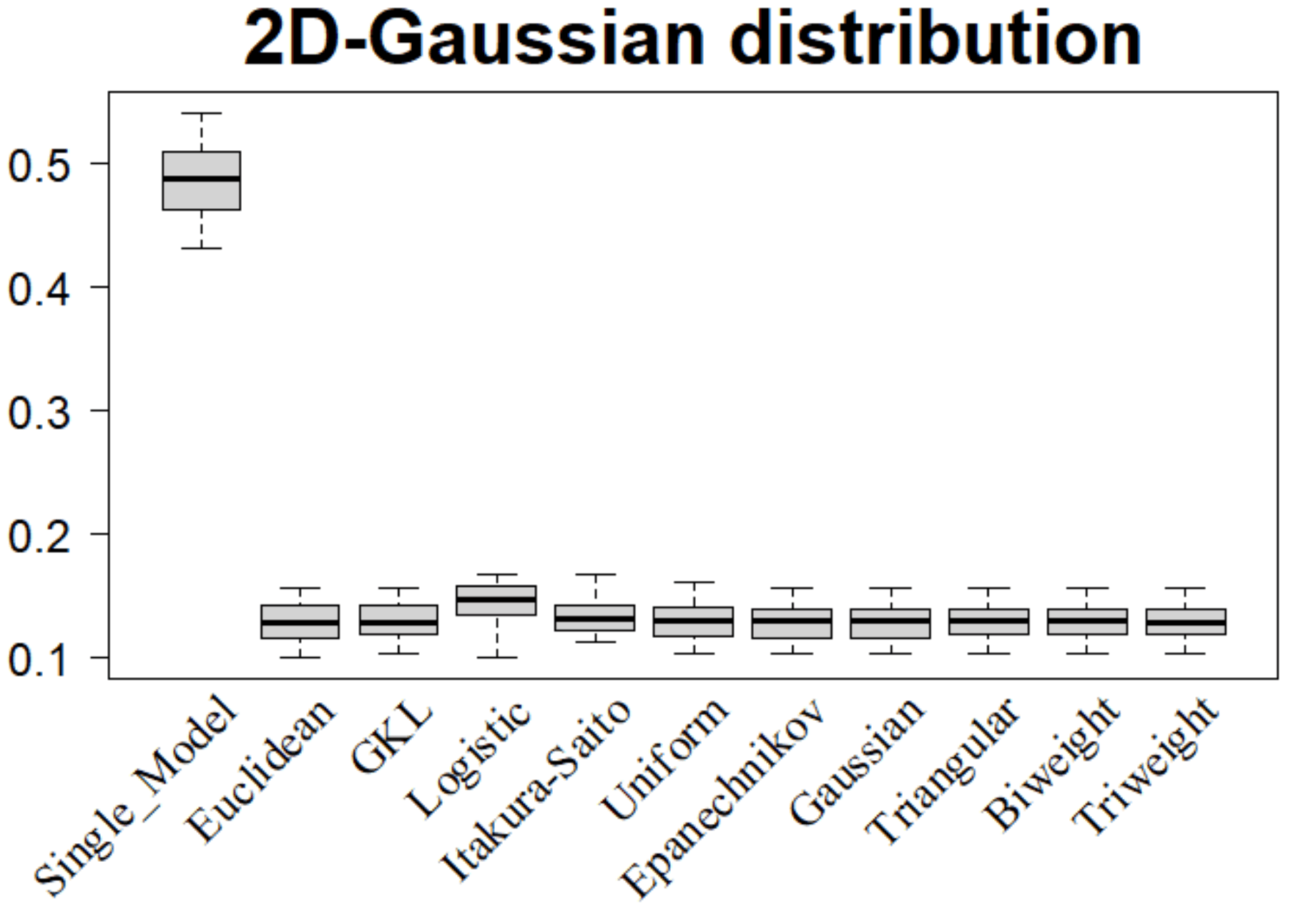}
\includegraphics[width = 6.5cm, height = 4.6cm]{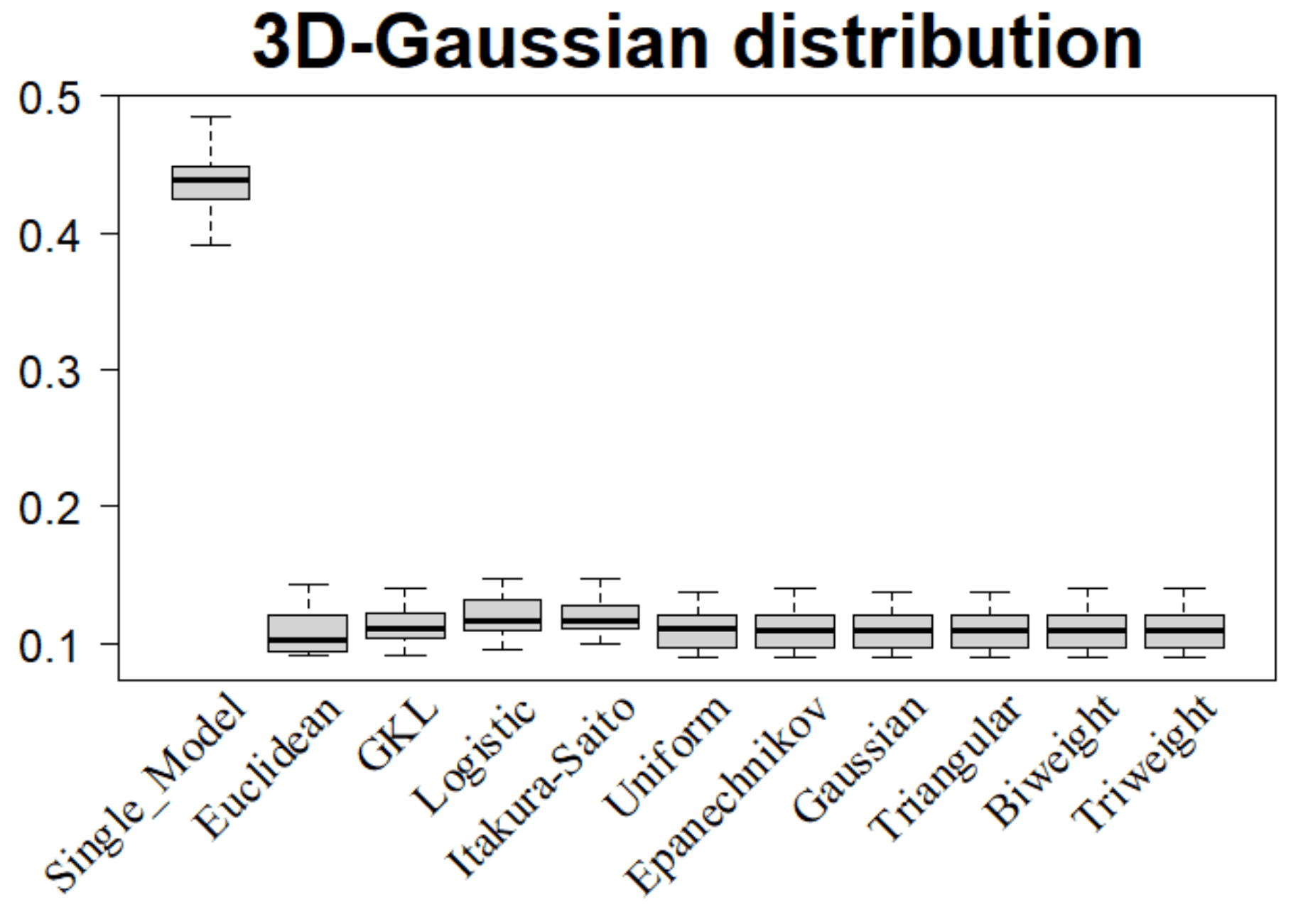}
\caption{Boxplots of misclassification error of $Comb_2^C$.}
\label{fig:5}
\end{figure}
The combined classification methods perform generally better than or similarly to the best individual estimator. The results of $Comb_3^C$, in the second row, seem to be better compared to the ones of $Comb_2^C$, in the first row, with remarkably smaller variances. We also note that the Gaussian kernel seems to be the most outstanding one among all kernel-based methods. \autoref{fig:5} and \autoref{fig:6} represent the boxplots of the associated average misclassification errors for $Comb_2^C$ and $Comb_3^C$ respectively (the results of the \autoref{tab:5})

\begin{sidewaystable}[ph!]
\small
\centering 
\begin{tabular}{l | c | c c c c | c c c c c c} 
\hline\hline            
\multirow{2}{*}{\bf Distribution} & \multirow{2}{*}{\bf Single} & \multicolumn{4}{c|}{{\bf Bregman divergence}} & \multicolumn{6}{c}{\textbf{Kernel}}\\ [0.5ex] \cline{3-12}
 & &Euclid & GKL & Logit & Ita & Unif & Epan & Gaus & Triang & Bi-wgt & Tri-wgt\\[0.5ex] 
\hline 
\multirow{4}{*}{Exp} & \multirow{2}{*}{\raisebox{-2.5ex}{$18.86$}} & \multirow{2}{*}{\raisebox{-2.5ex}{$8.58$}} & \multirow{2}{*}{\raisebox{-2.5ex}{$7.42$}} & \multirow{2}{*}{\raisebox{-2.5ex}{$4.09$}} & $\multirow{2}{*}{\raisebox{-2ex}{\textbf{3.92}}}$ & $3.49$ & $3.51$ & $\textcolor{blue}{\textbf{3.46}}$ & $3.51$ & $3.56$ & $3.56$ \\
 & & & & & & $(0.89)$ & $(0.94)$ & $(0.88)$ & $(0.94)$ & $(0.91)$ & $(0.91)$ \\ \cline{7-12}
 & \multirow{2}{*}{\raisebox{2.5ex}{$(1.70)$}} & \multirow{2}{*}{\raisebox{2.5ex}{$(1.77)$}} & \multirow{2}{*}{\raisebox{2.5ex}{$(1.55)$}} & \multirow{2}{*}{\raisebox{2.5ex}{$(1.08)$}} & \multirow{2}{*}{\raisebox{2.5ex}{$(1.15)$}} & $2.91$ & $2.63$ & $2.49$ & $2.70$ & $2.56$ & $\textcolor{red}{\textbf{2.46}}$ \\
 & & & & & & $(0.81)$ & $(0.70)$ & $(0.74)$ & $(0.75)$ & $(0.63)$ & $(0.66)$ \\
 \hline             
 \multirow{4}{*}{Pois} & \multirow{2}{*}{\raisebox{-2.5ex}{$46.93$}} & \multirow{2}{*}{\raisebox{-2.5ex}{$9.19$}} & \multirow{2}{*}{\raisebox{-2.5ex}{$\textbf{8.45}$}} & \multirow{2}{*}{\raisebox{-2.5ex}{$13.33$}} & $\multirow{2}{*}{\raisebox{-2ex}{10.15}}$ & $8.59$ & $\textcolor{blue}{\textbf{8.51}}$ & $\textcolor{blue}{\textbf{8.51}}$ & $\textcolor{blue}{\textbf{8.51}}$ & $8.52$ & $8.52$ \\
 & & & & & & $(1.37)$ & $(1.46)$ & $(1.47)$ & $(1.46)$ & $(1.47)$ & $(1.49)$ \\ \cline{7-12}
 & \multirow{2}{*}{\raisebox{2.5ex}{$(3.35)$}} & \multirow{2}{*}{\raisebox{2.5ex}{$(1.27)$}} & \multirow{2}{*}{\raisebox{2.5ex}{$(1.24)$}} & \multirow{2}{*}{\raisebox{2.5ex}{$(1.84)$}} & \multirow{2}{*}{\raisebox{2.5ex}{$(1.47)$}} & $8.51$ & $8.46$ & $8.44$ & $\textcolor{red}{\textbf{8.42}}$ & $8.57$ & $8.44$ \\
 & & & & & & $(1.28)$ & $(1.11)$ & $(1.17)$ & $(1.15)$ & $(1.28)$ & $(1.13)$ \\
 \hline             
\multirow{4}{*}{Geom} & \multirow{2}{*}{\raisebox{-2.5ex}{$19.90 $}} & \multirow{2}{*}{\raisebox{-2.5ex}{$12.57$}} & \multirow{2}{*}{\raisebox{-2.5ex}{$4.71$}} & \multirow{2}{*}{\raisebox{-2.5ex}{$\textbf{3.94}$}} & $\multirow{2}{*}{\raisebox{-2ex}{8.12}}$ & $3.61$ & $\textcolor{blue}{\textbf{3.60}}$ & $\textcolor{blue}{\textbf{3.60}}$ & $3.61$ & $\textcolor{blue}{\textbf{3.60}}$ & $\textcolor{blue}{\textbf{3.60}}$ \\
 & & & & & & $(1.15)$ & $(1.16)$ & $(1.16)$ & $(1.15)$ & $(1.16)$ & $(1.16)$ \\ \cline{7-12}
 & \multirow{2}{*}{\raisebox{2.5ex}{$(2.07)$}} & \multirow{2}{*}{\raisebox{2.5ex}{$(2.39)$}} & \multirow{2}{*}{\raisebox{2.5ex}{$(2.37)$}} & \multirow{2}{*}{\raisebox{2.5ex}{$(1.15)$}} & \multirow{2}{*}{\raisebox{2.5ex}{$(1.57)$}} & $3.76$ & $3.52$ & $\textcolor{red}{\textbf{2.94}}$ & $3.48$ & $3.47$ & $3.40$ \\
 & & & & & & $(0.92)$ & $(1.11)$ & $(0.93)$ & $(1.09)$ & $(1.11)$ & $(1.06)$ \\
 \hline             
\multirow{4}{*}{2D Gaus} & \multirow{2}{*}{\raisebox{-2.5ex}{$49.00 $}} & \multirow{2}{*}{\raisebox{-2.5ex}{$\textbf{12.37}$}} & \multirow{2}{*}{\raisebox{-2.5ex}{$12.40$}} & \multirow{2}{*}{\raisebox{-2.5ex}{$14.14$}} & $\multirow{2}{*}{\raisebox{-2ex}{13.05}}$ & $12.87$ & $12.82$ & $\textcolor{blue}{\textbf{12.80}}$ & $12.84$ & $12.84$ & $12.87$ \\
 & & & & & & $(1.60 )$ & $(1.59 )$ & $(1.56 )$ & $(1.57)$ & $(1.57)$ & $(1.60)$ \\ \cline{7-12}
 & \multirow{2}{*}{\raisebox{2.5ex}{$(2.52)$}} & \multirow{2}{*}{\raisebox{2.5ex}{$(1.55)$}} & \multirow{2}{*}{\raisebox{2.5ex}{$(1.50)$}} & \multirow{2}{*}{\raisebox{2.5ex}{$(1.44)$}} & \multirow{2}{*}{\raisebox{2.5ex}{$(1.61)$}} & $\textcolor{red}{\textbf{12.02}}$ & $12.11$ & $12.06$ & $12.11$ & $12.09$ & $12.10$ \\
 & & & & & & $(1.30)$ & $(1.24)$ & $(1.35)$ & $(1.27)$ & $(1.23)$ & $(1.22)$ \\
 \hline              
\multirow{4}{*}{3D Gaus} & \multirow{2}{*}{\raisebox{-2.5ex}{$43.39 $}} & \multirow{2}{*}{\raisebox{-2.5ex}{$\textbf{10.77}$}} & \multirow{2}{*}{\raisebox{-2.5ex}{$10.99$}} & \multirow{2}{*}{\raisebox{-2.5ex}{$11.74$}} & $\multirow{2}{*}{\raisebox{-2ex}{11.56}}$ & $11.08$ & $11.01$ & $\textcolor{blue}{\textbf{11.00}}$ & $\textcolor{blue}{\textbf{11.00}}$ & $11.04$ & $11.03$ \\
 & & & & & & $(1.58)$ & $(1.52)$ & $(1.50)$ & $(1.50)$ & $(1.57)$ & $(1.55)$ \\ \cline{7-12}
 & \multirow{2}{*}{\raisebox{2.5ex}{$(2.52)$}} & \multirow{2}{*}{\raisebox{2.5ex}{$(1.40)$}} & \multirow{2}{*}{\raisebox{2.5ex}{$(1.44)$}} & \multirow{2}{*}{\raisebox{2.5ex}{$(1.45)$}} & \multirow{2}{*}{\raisebox{2.5ex}{$(1.51)$}} & $10.23$ & $9.93$ & $\textcolor{red}{\textbf{9.76}}$ & $10.04$ & $9.83$ & $9.84$ \\
 & & & & & & $(1.40)$ & $(1.47)$ & $(1.53)$ & $(1.47)$ & $(1.61)$ & $(1.61)$ \\
 \hline             
\end{tabular}
\caption{Misclassification errors of $Comb_2^C$ and $Comb_3^C$ computed over $20$ runs of all simulated data (1 unit = $10^{-2}$).}
\label{tab:5}
\end{sidewaystable}

\newpage
\subsubsection{Regression}
\label{subsec:reg}
In the regression case, the results in the \autoref{tab:6} again agree with the NMI results given in \autoref{tab:4}, except for Geometric distribution, where the estimator based on Generalized Kullback-Leibler Divergence outperforms the estimator built after clustering with Logistic divergence. Again, the performance of the estimators is globally improved by combining. It is clear that Gaussian kernel does the best job, and $Comb_2^R$ and $Comb_3^R$ alternatively outperform each other.

\begin{figure}[ph!]
\centering
\includegraphics[width = 6.2cm, height = 4.1cm]{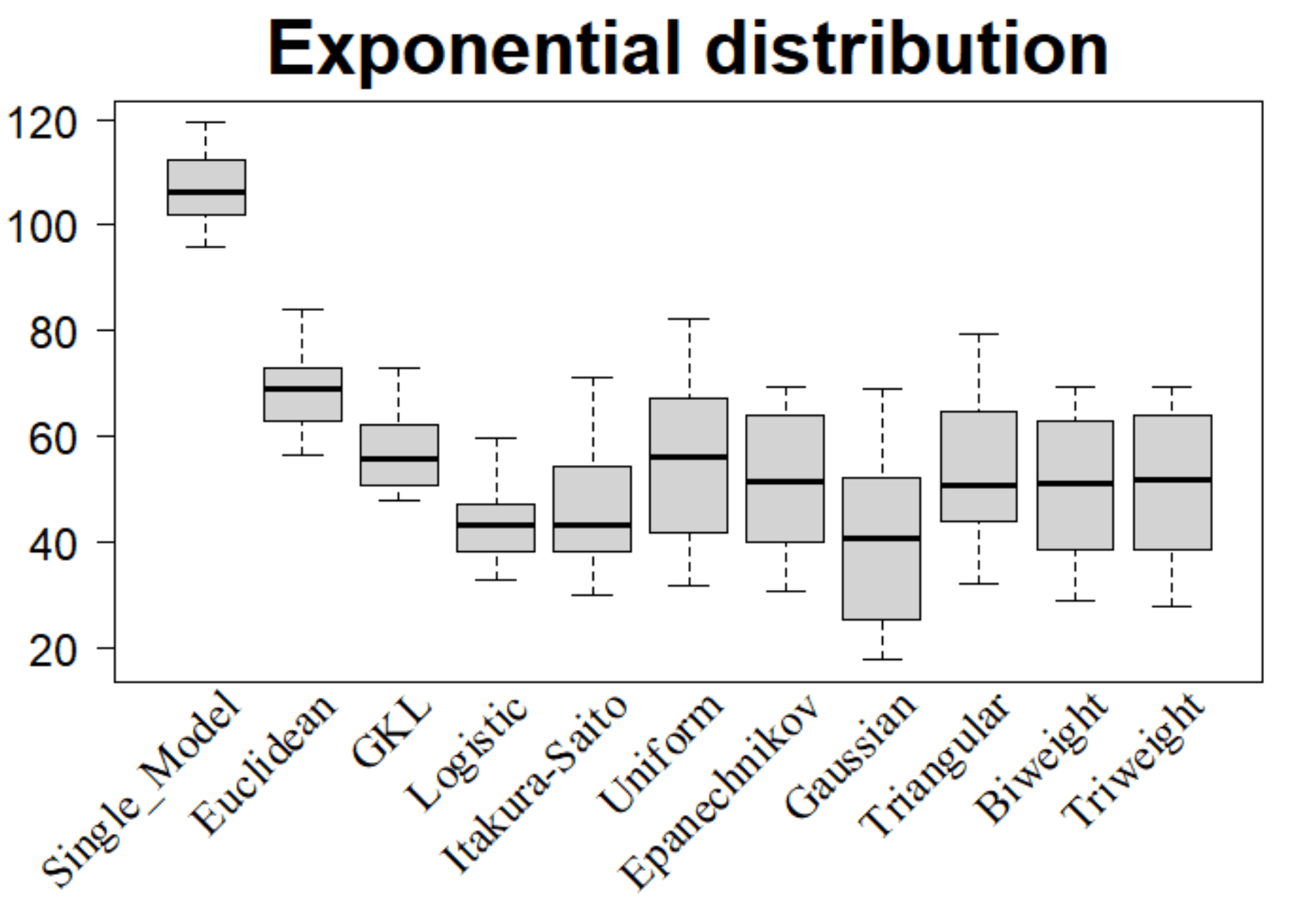}
\includegraphics[width = 6.2cm, height = 4.1cm]{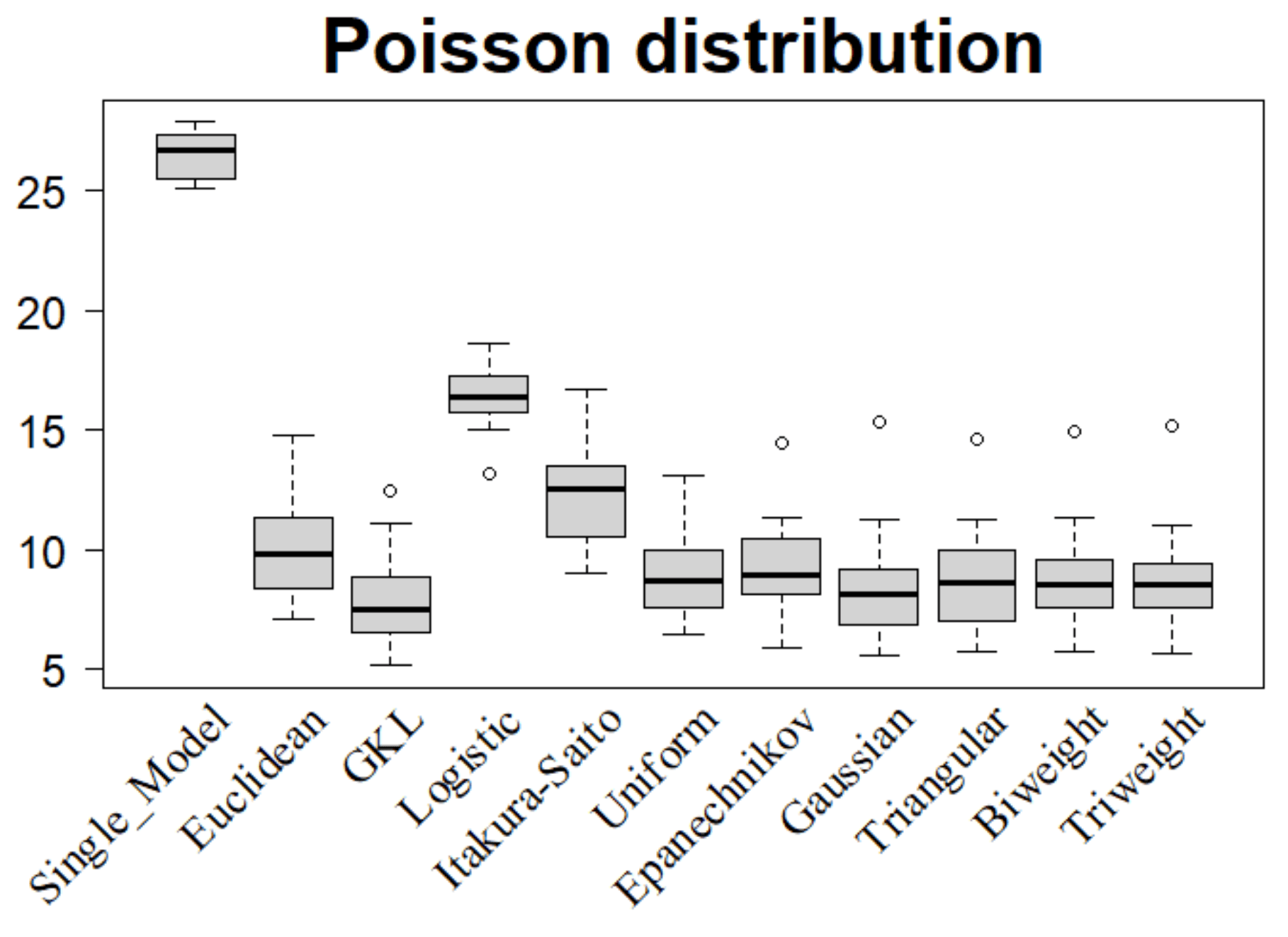}
\includegraphics[width = 6.2cm, height = 4.1cm]{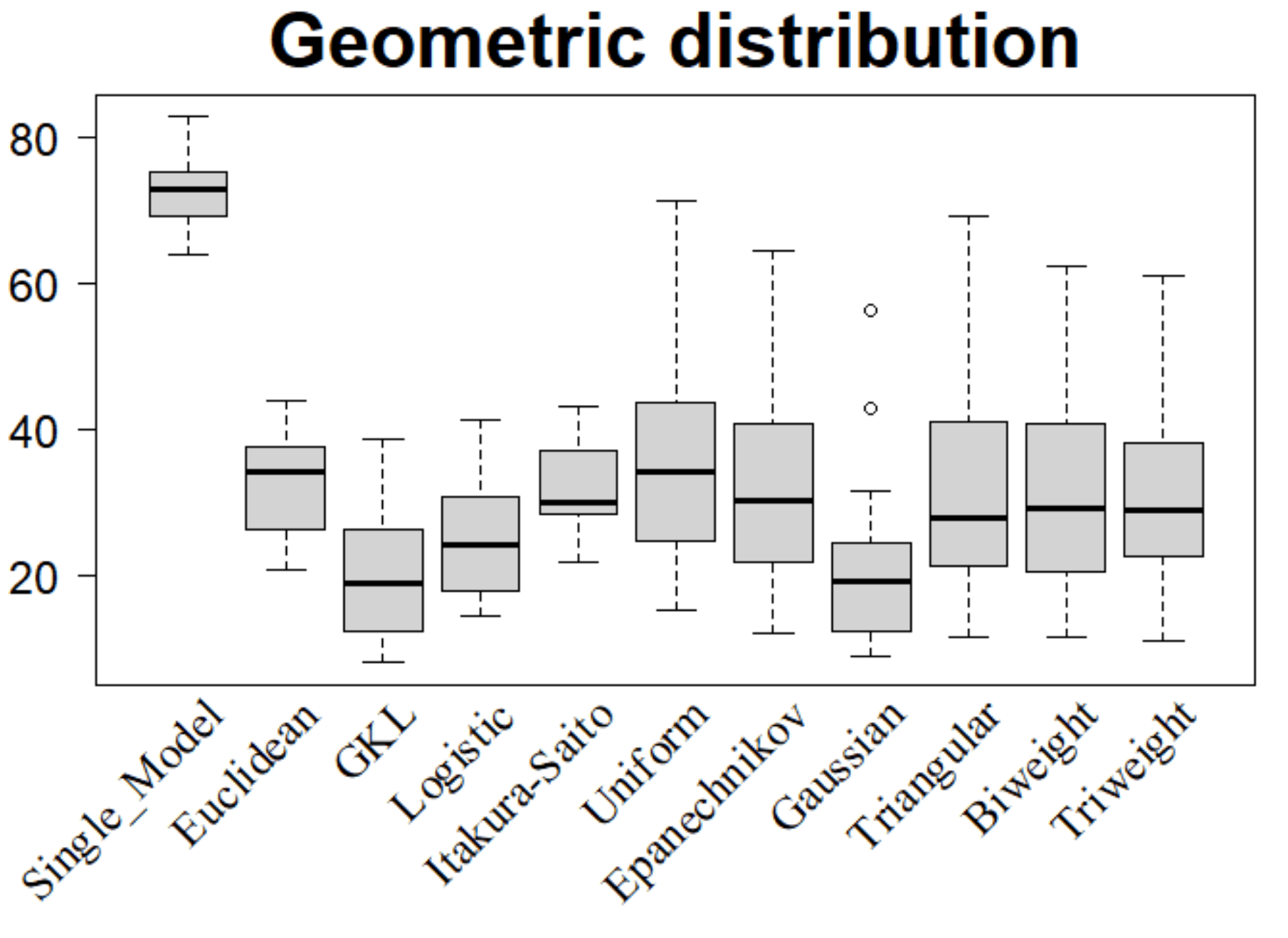}
\includegraphics[width = 6.2cm, height = 4.1cm]{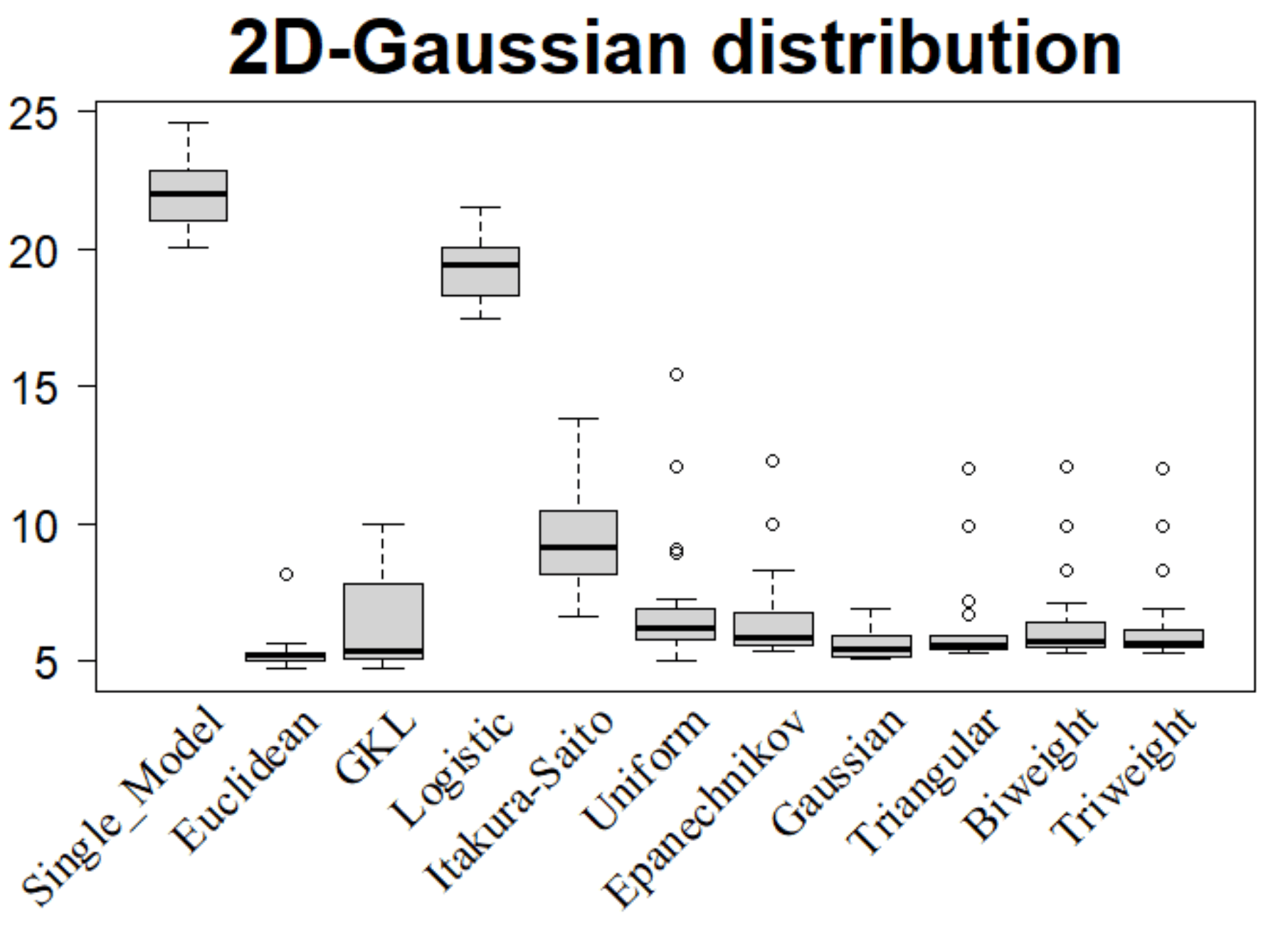}
\includegraphics[width = 6.2cm, height = 4.1cm]{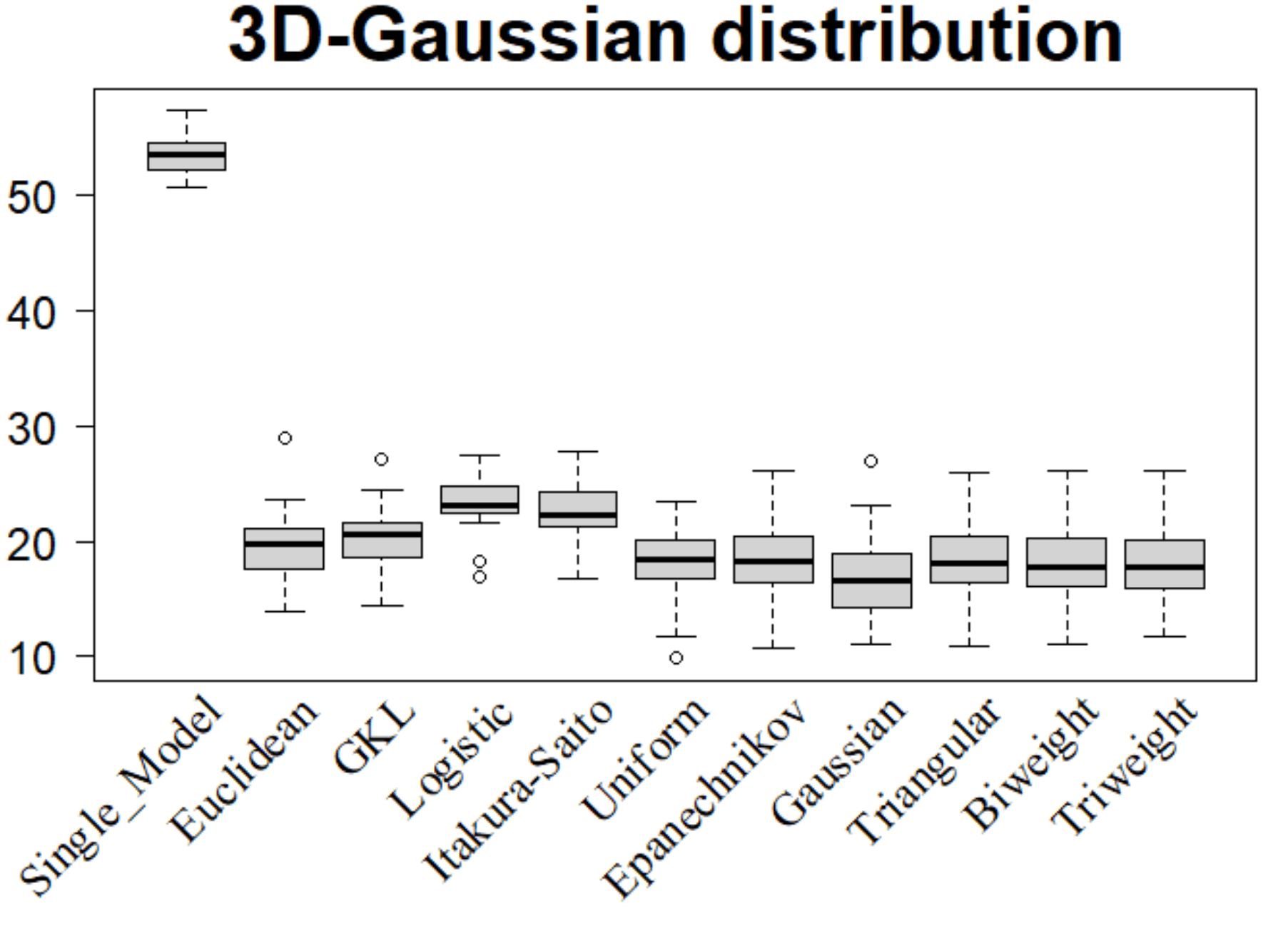}
\caption{Boxplots of RMSE of $Comb_2^R$.}
\label{fig:7}
\end{figure}

\newpage
\begin{figure}[ph!]
\centering
\includegraphics[width = 6.2cm, height = 4.2cm]{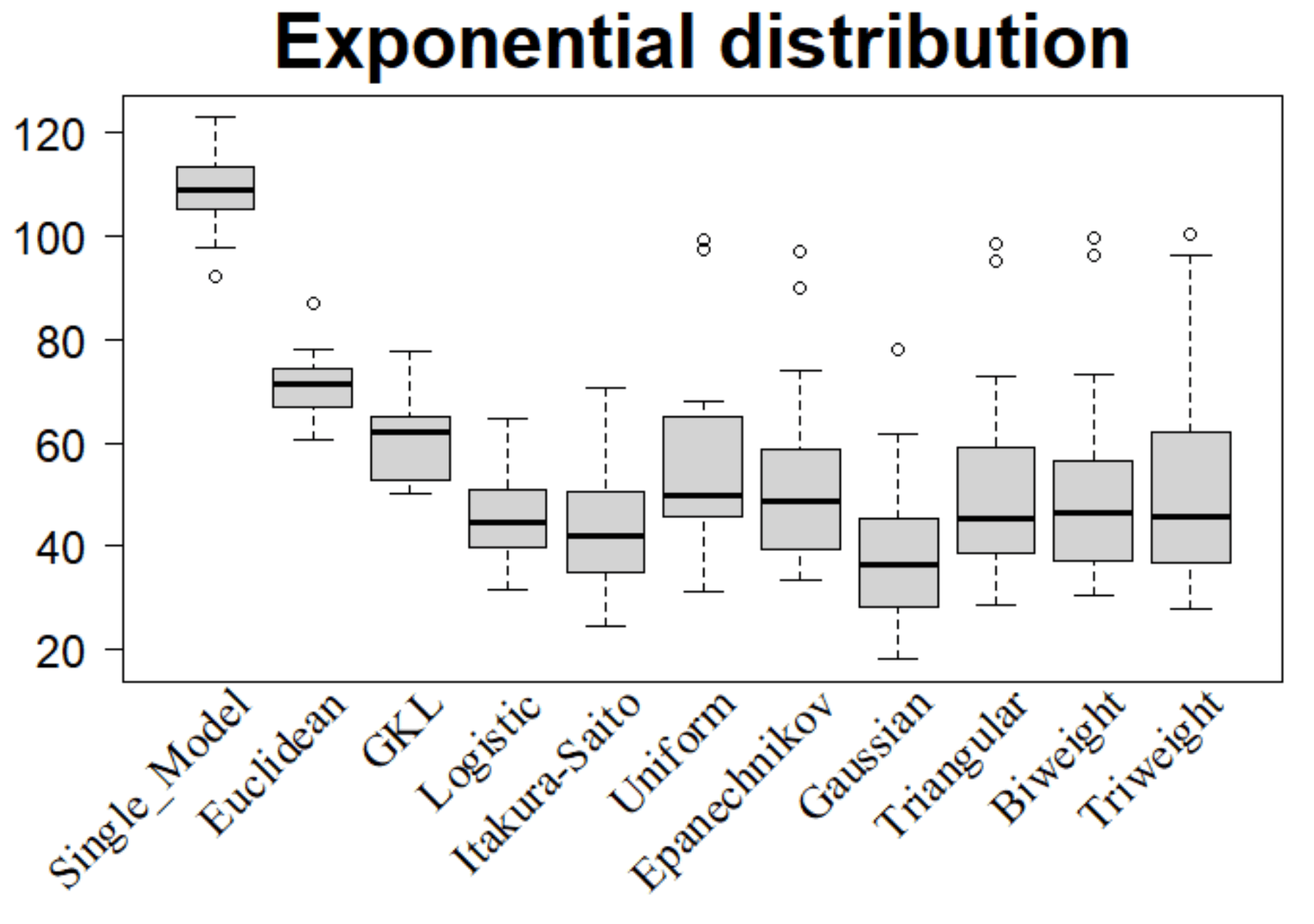}
\includegraphics[width = 6.2cm, height = 4.2cm]{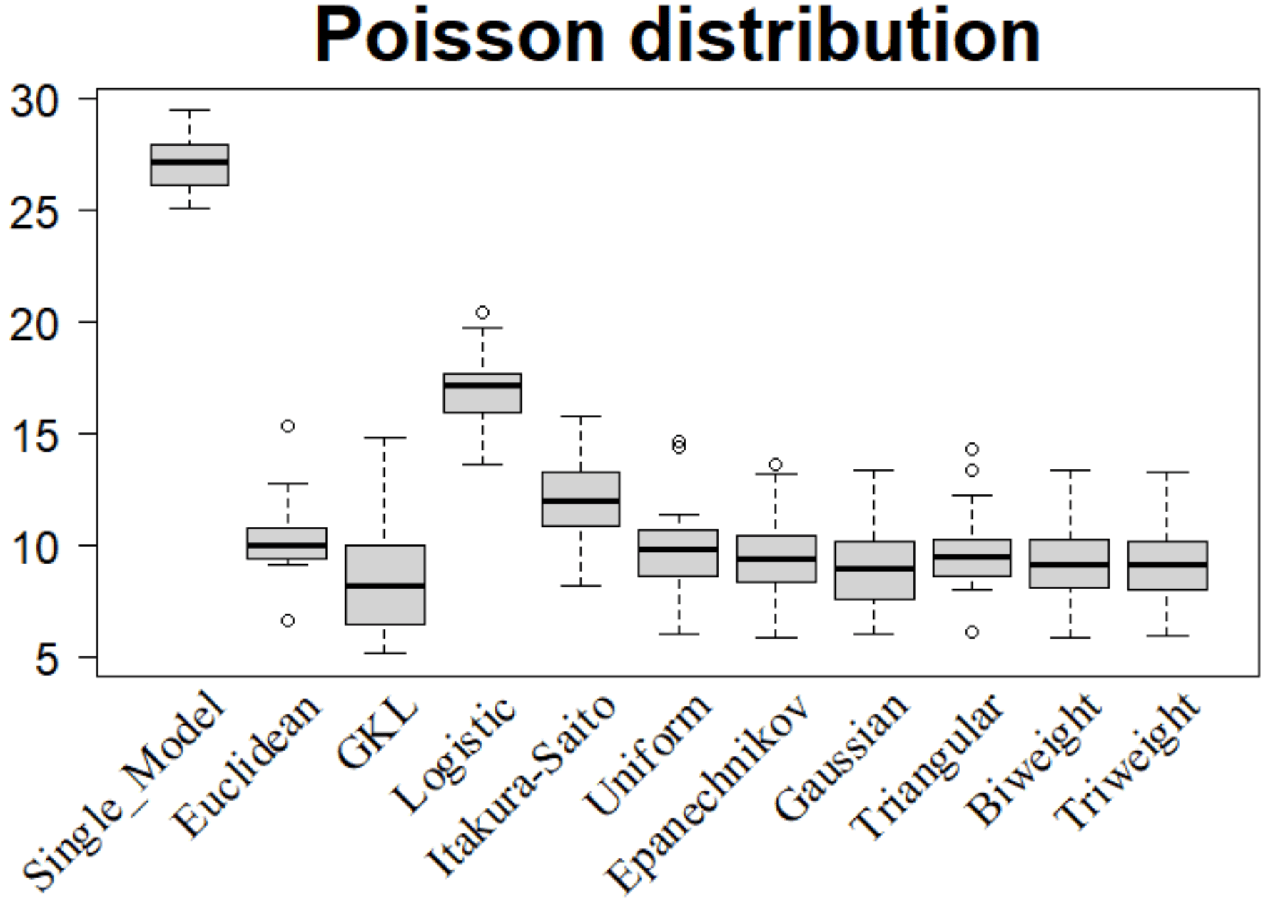}
\includegraphics[width = 6.2cm, height = 4.2cm]{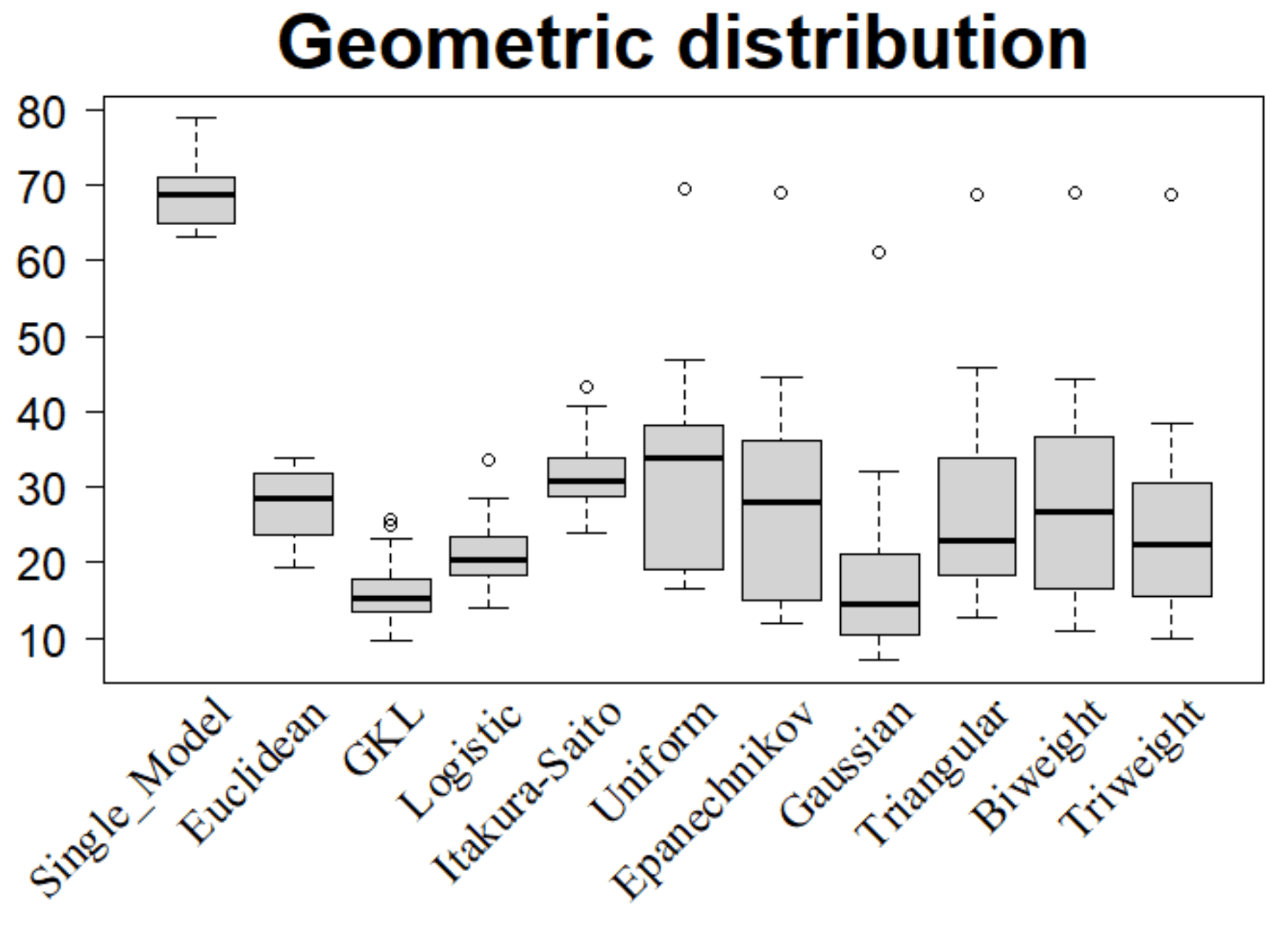}
\includegraphics[width = 6.2cm, height = 4.2cm]{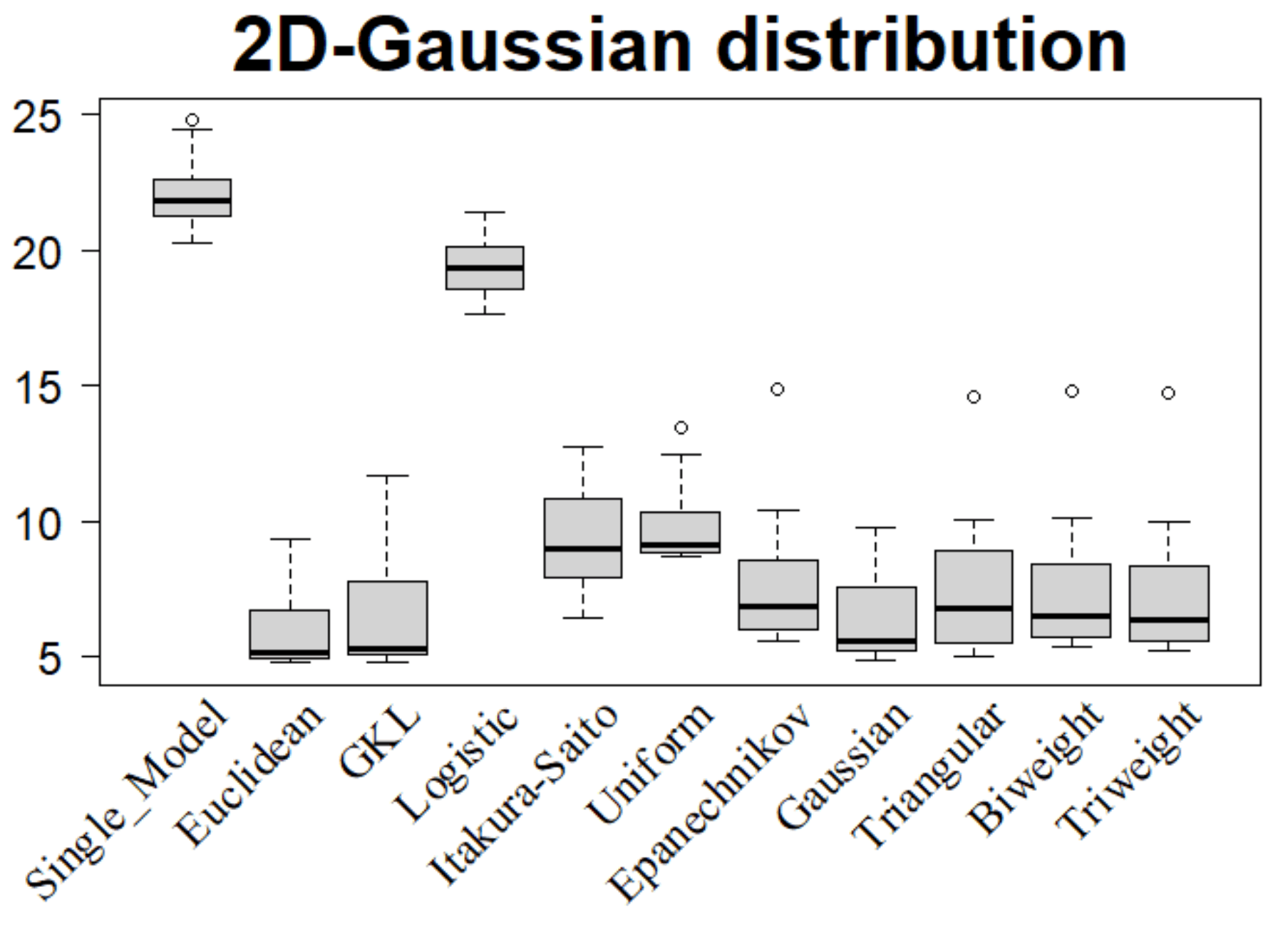}
\includegraphics[width = 6.2cm, height = 4.2cm]{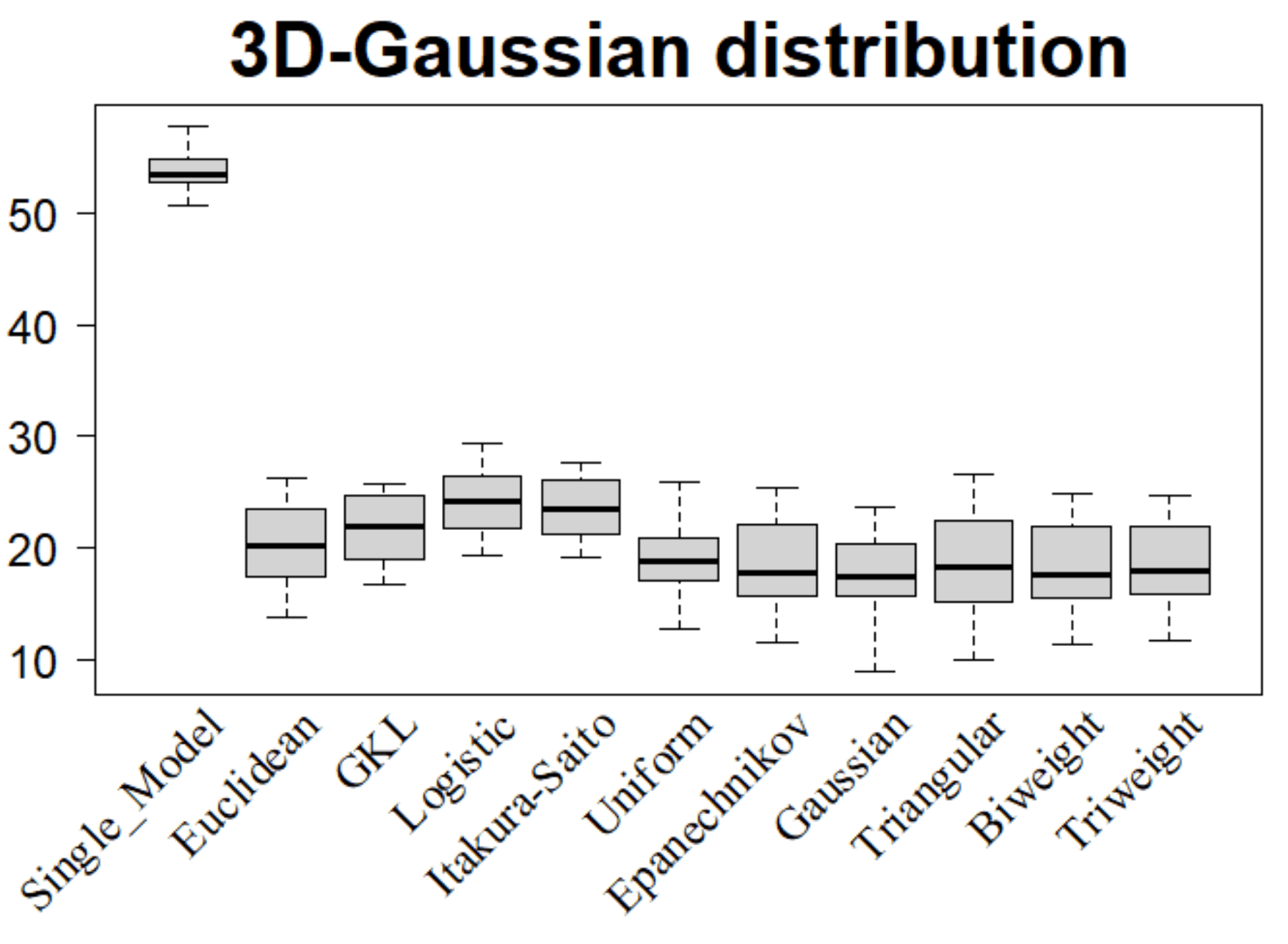}
\caption{Boxplots of RMSE of $Comb_3^R$.}
\label{fig:8}
\end{figure}

\autoref{fig:7} and \autoref{fig:8} above represent the associated boxplots of root mean square errors for $Comb_2^R$ and $Comb_3^R$ respectively (the results of the \autoref{tab:6}). 

\begin{sidewaystable}[ph!]
\small
\centering 
\begin{tabular}{l | c | c c c c | c c c c c c}
\hline\hline            
\multirow{2}{*}{\bf Distribution} & \multirow{2}{*}{\bf Single} & \multicolumn{4}{c|}{{\bf Bregman divergence}} & \multicolumn{6}{c}{\textbf{Kernel}}\\ [0.5ex] \cline{3-12}
 & &Euclid & GKL & Logit & Ita & Unif & Epan & Gaus & Triang & Bi-wgt & Tri-wgt\\[0.5ex]
\hline 
\multirow{4}{*}{Exp} & \multirow{2}{*}{\raisebox{-2.5ex}{$106.58$}} & \multirow{2}{*}{\raisebox{-2.5ex}{$68.74$}} & \multirow{2}{*}{\raisebox{-2.5ex}{$57.06$}} & \multirow{2}{*}{\raisebox{-2.5ex}{$44.54$}} & $\multirow{2}{*}{\raisebox{-2ex}{\textbf{44.46}}}$ & $55.11$ & $51.14$ & $\textcolor{blue}{\textbf{40.21}}$ & $52.99$ & $50.24$ & $50.64$ \\
 & & & & & & $(15.85)$ & $(13.31)$ & $(14.40)$ & $(13.12)$ & $(13.74)$ & $(14.41)$ \\ \cline{7-12}
 & \multirow{2}{*}{\raisebox{2.5ex}{$(7.13)$}} & \multirow{2}{*}{\raisebox{2.5ex}{$(6.84)$}} & \multirow{2}{*}{\raisebox{2.5ex}{$(7.37)$}} & \multirow{2}{*}{\raisebox{2.5ex}{$(7.37)$}} & \multirow{2}{*}{\raisebox{2.5ex}{$(10.96)$}} & $56.34$ & $52.62$ & $\textcolor{red}{\textbf{39.12}}$ & $51.31$ & $51.20$ & $51.98$ \\
 & & & & & & $(17.48)$ & $(17.82)$ & $(14.98)$ & $(19.55)$ & $(19.69)$ & $(20.12)$ \\
 \hline             
 \multirow{4}{*}{Pois} & \multirow{2}{*}{\raisebox{-2.5ex}{$26.76$}} & \multirow{2}{*}{\raisebox{-2.5ex}{$10.16$}} & \multirow{2}{*}{\raisebox{-2.5ex}{$\textbf{8.22}$}} & \multirow{2}{*}{\raisebox{-2.5ex}{$16.72$}} & $\multirow{2}{*}{\raisebox{-2ex}{12.15}}$ & $8.88$ & $9.18$ & $\textcolor{blue}{\textbf{8.43}}$ & $8.85$ & $8.84$ & $8.76$ \\
 & & & & & & $(1.65)$ & $(1.98)$ & $(2.18)$ & $(2.06)$ & $(2.03)$ & $(2.03)$ \\ \cline{7-12}
 & \multirow{2}{*}{\raisebox{2.5ex}{$(1.11)$}} & \multirow{2}{*}{\raisebox{2.5ex}{$(1.91)$}} & \multirow{2}{*}{\raisebox{2.5ex}{$(2.25)$}} & \multirow{2}{*}{\raisebox{2.5ex}{$(1.61)$}} & \multirow{2}{*}{\raisebox{2.5ex}{$(1.86)$}} & $9.73$ & $9.61$ & $\textcolor{red}{\textbf{9.13}}$ & $9.64$ & $9.40$ & $9.43$ \\
 & & & & & & $(2.25)$ & $(1.86)$ & $(1.92)$ & $(1.91)$ & $(1.86)$ & $(1.93)$ \\
 \hline             
\multirow{4}{*}{Geom} & \multirow{2}{*}{\raisebox{-2.5ex}{$70.45$}} & \multirow{2}{*}{\raisebox{-2.5ex}{$29.99$}} & \multirow{2}{*}{\raisebox{-2.5ex}{$\textbf{18.33}$}} & \multirow{2}{*}{\raisebox{-2.5ex}{$22.94$}} & $\multirow{2}{*}{\raisebox{-2ex}{31.94}}$ & $36.39$ & $32.49$ & $\textcolor{blue}{\textbf{21.51}}$ & $31.48$ & $31.44$ & $30.89$ \\
 & & & & & & $(13.81)$ & $(13.49)$ & $(11.79)$ & $(14.31)$ & $(13.51)$ & $(12.21)$ \\ \cline{7-12}
 & \multirow{2}{*}{\raisebox{2.5ex}{$(4.52)$}} & \multirow{2}{*}{\raisebox{2.5ex}{$(5.95)$}} & \multirow{2}{*}{\raisebox{2.5ex}{$(7.34)$}} & \multirow{2}{*}{\raisebox{2.5ex}{$(6.21)$}} & \multirow{2}{*}{\raisebox{2.5ex}{$(5.19)$}} & $31.83$ & $27.90$ & $\textcolor{red}{\textbf{17.82}}$ & $26.82$ & $28.45$ & $24.58$ \\
 & & & & & & $(12.88)$ & $(14.20)$ & $(12.58)$ & $(13.28)$ & $(14.02)$ & $(13.21)$ \\
 \hline             
\multirow{4}{*}{2D Gaus} & \multirow{2}{*}{\raisebox{-2.5ex}{$21.98$}} & \multirow{2}{*}{\raisebox{-2.5ex}{$\textbf{5.63}$}} & \multirow{2}{*}{\raisebox{-2.5ex}{$6.46$}} & \multirow{2}{*}{\raisebox{-2.5ex}{$19.36$}} & $\multirow{2}{*}{\raisebox{-2ex}{9.38}}$ & $7.09$ & $6.57$ & $\textcolor{blue}{\textbf{5.57}}$ & $6.20$ & $6.41$ & $6.33$ \\
 & & & & & & $(2.55)$ & $(1.78)$ & $(0.49)$ & $(1.72)$ & $(1.76)$ & $(1.75)$ \\ \cline{7-12}
 & \multirow{2}{*}{\raisebox{2.5ex}{$(1.20)$}} & \multirow{2}{*}{\raisebox{2.5ex}{$(1.26)$}} & \multirow{2}{*}{\raisebox{2.5ex}{$(1.81)$}} & \multirow{2}{*}{\raisebox{2.5ex}{$(1.11)$}} & \multirow{2}{*}{\raisebox{2.5ex}{$(1.86)$}} & $9.75$ & $7.70$ & $\textcolor{red}{\textbf{6.42}}$ & $7.45$ & $7.47$ & $7.34$ \\
 & & & & & & $(1.30)$ & $(2.24)$ & $(1.49)$ & $(2.42)$ & $(2.28)$ & $(2.31)$ \\
 \hline              
\multirow{4}{*}{3D Gaus} & \multirow{2}{*}{\raisebox{-2.5ex}{$53.55$}} & \multirow{2}{*}{\raisebox{-2.5ex}{$\textbf{19.89}$}} & \multirow{2}{*}{\raisebox{-2.5ex}{$20.93$}} & \multirow{2}{*}{\raisebox{-2.5ex}{$23.71$}} & $\multirow{2}{*}{\raisebox{-2ex}{22.96}}$ & $18.16$ & $18.20$ & $\textcolor{blue}{\textbf{16.94}}$ & $18.25$ & $18.05$ & $18.00$ \\
 & & & & & & $3.42)$ & $(3.45)$ & $(4.06)$ & $(3.41)$ & $(3.50)$ & $(3.49)$ \\ \cline{7-12}
 & \multirow{2}{*}{\raisebox{2.5ex}{$(1.74)$}} & \multirow{2}{*}{\raisebox{2.5ex}{$(3.49)$}} & \multirow{2}{*}{\raisebox{2.5ex}{$(2.97)$}} & \multirow{2}{*}{\raisebox{2.5ex}{$(2.70)$}} & \multirow{2}{*}{\raisebox{2.5ex}{$(2.74)$}} & $19.24$ & $18.52$ & $\textcolor{red}{\textbf{17.51}}$ & $18.64$ & $18.19$ & $18.42$ \\
 & & & & & & $(3.54)$ & $(4.02)$ & $(3.64)$ & $(4.37)$ & $(3.91)$ & $(3.68)$ \\
 \hline             
\end{tabular}
\caption{RMSE of $Comb_2^R$ and $Comb_3^R$ computed over $20$ runs of all simulated data.}
\label{tab:6}
\end{sidewaystable}

The numerical results are quite satisfactory, and this is a piece of evidence showing that KFC procedure is an interesting method for building predictive models, especially when the number of existing groups of the input data is available. It is even more interesting in the next section where the procedure is implemented on a real dataset of Air compressor machine for which the number of clustering is not available.

Throughout the simulation, we could see that the procedure is time-consuming, especially when the implementation is done with more options of Bregman divergences. However, it should be pointed out that the structure of KFC procedure is parallel in a sense that the {\it K} and {\it F} steps ({\it $K$-means} and {\it Fit} step) of the procedure can be implemented in parallel independently, and only the predictions given by all of those independently constructed estimators are required in the consensual aggregation step.

\section{Application}
\label{section:appli}
{
In this section, we study the performance of the KFC procedure on real data.
The goal of the application here is to model the power consumption of an air compressor equipment \cite{CHM}.
The target is the electrical power of the machine, and 6 explanatory variables are available:
 air temperature, input pressure, output pressure, flow, water temperature.
 The dataset contains $N = 2000$ hourly observations of a working air compressor.
We run the algorithms over $20$ random partitions of $80\%$ training sample. The root mean square error (RMSE) computed on the testing sets as well as the associated standard errors are summarized in \autoref{tab:7}. As the number of clusters is unknown, we perform the KFC algorithm with different values of the number of clusters $K\in\{1,2,...,8\}$. For the consensual aggregation step, we use a Gaussian kernel which showed to be the best one in the simulations with synthetic data. Note that for the simple linear model with only one cluster on the whole dataset ($K=1$), we obtain the average RMSE of ${\bf 178.67}$ with the associated standard error of ${\bf 5.47}$.}

The associated boxplots are given in \autoref{fig:9} below. We observe that the performance of the individual estimators improve as the number $K$ of clusters increases. Note that $Comb_3^R$ outperforms $Comb_2^R$ with much lower error {(reduced more than $20\%$ of error given by $Comb_2^R$)} and also a smaller variance. Regardless of the number of clusters, the combination step allows to reduce the RMSE in each case to approximately the same level. Hence, our strategy may be interesting even without the knowledge of the number of clusters. 

\begin{table}[h!]
\centering 
\begin{tabular}{c | c c c c c c} 
\hline\hline            
$K$ & Euclid & GKL & Logistic & Ita & $Comb_2^R$ & $Comb_3^R$
\\ [0.5ex]  
\hline    
\multirow{2}{*}{2} & $158.85$ & $158.90$ & $159.35$ & $158.96$ & $153.34$ & $\textcolor{blue}{\textbf{116.69}}$\\
& $(6.42)$ & $(6.48)$ & $(6.71)$ & $(6.41)$ & $(6.72)$ & $(5.86)$\\ 
\hline    
\multirow{2}{*}{3} & $157.38$ & $157.24$ & $156.99$ & $157.24$ & $153.69$ & $\textcolor{blue}{\textbf{117.45}}$\\
& $(6.95)$ & $(6.84)$ & $(6.65)$ & $(6.85)$ & $(6.64)$ & $(5.55)$\\ 
\hline    
\multirow{2}{*}{4} & $154.33$ & $153.96$ & $153.99$ & $154.07$ & $152.09$ & $\textcolor{blue}{\textbf{117.16}}$\\
& $(6.69)$ & $(6.74)$ & $(6.45)$ & $(7.01)$ & $(6.58)$ & $(5.99)$\\ 
\hline    
\multirow{2}{*}{5} & $153.18$ & $153.19$ & $152.95$ & $152.25$ & $151.05$ & $\textcolor{blue}{\textbf{117.55}}$\\
& $(6.91)$ & $(6.77)$ & $(6.57)$ & $(6.70)$ & $(6.76)$ & $(5.90)$\\
\hline    
\multirow{2}{*}{6} & $151.16$ & $151.67$ & $151.89$ & $151.75$ & $150.27$ & $\textcolor{blue}{\textbf{117.74}}$\\
& $(6.91)$ & $(6.96)$ & $(6.62)$ & $(6.57)$ & $(6.82)$ & $(5.86)$\\ 
\hline    
\multirow{2}{*}{7} & $151.08$ & $150.99$ & $152.81$ & $151.85$ & $150.46$ & $\textcolor{blue}{\textbf{117.58}}$\\
& $(6.77)$ & $(6.84)$ & $(7.11)$ & $(6.61)$ & $(6.87)$ & $(6.15)$\\
\hline    
\multirow{2}{*}{8} & $151.27$ & $151.09$ & $152.07$ & $150.90$ & $150.21$ & $\textcolor{blue}{\textbf{117.91}}$\\
& $(7.17)$ & $(7.01)$ & $(6.65)$ & $(6.96)$ &$(7.03)$ & $(5.83)$\\
\hline             
\end{tabular}
\caption{Average RMSE of each algorithm performed on Air Compressor data.}%
\label{tab:7}
\end{table}

\begin{figure}[ph!]
\centering
\includegraphics[width = 6.3cm, height = 4.2cm]{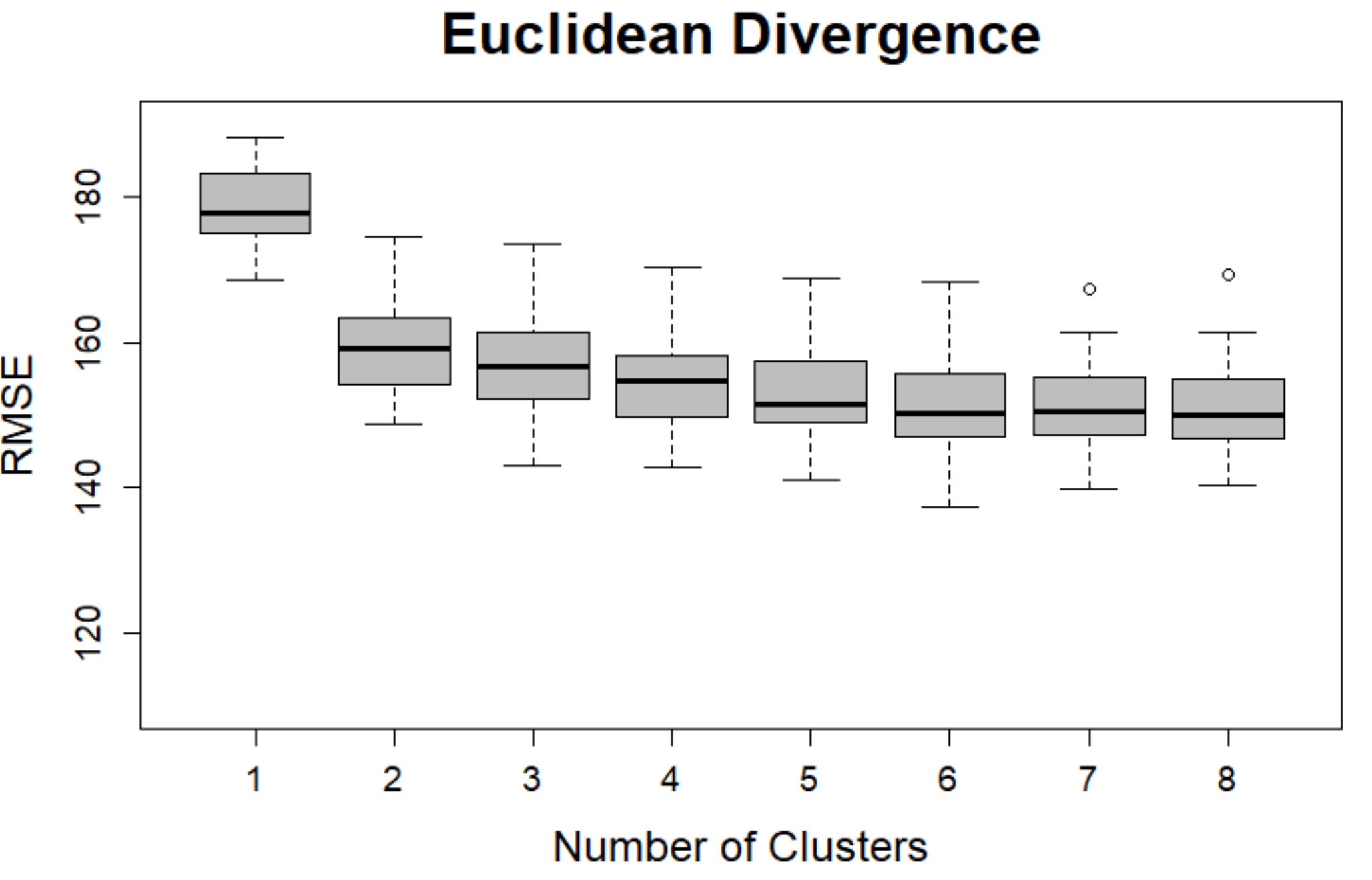}
\includegraphics[width = 6.3cm, height = 4.2cm]{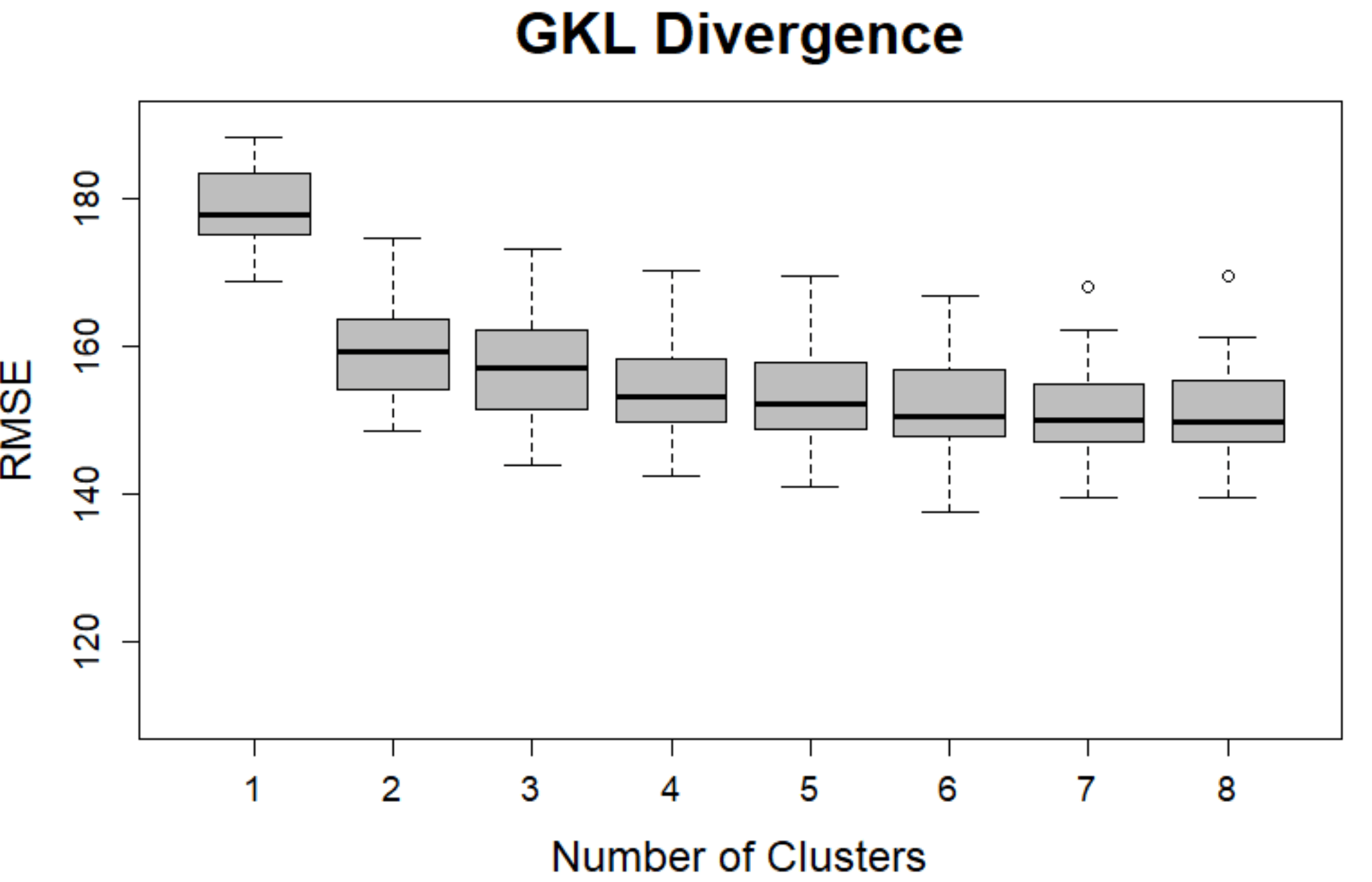}
\includegraphics[width = 6.3cm, height = 4.2cm]{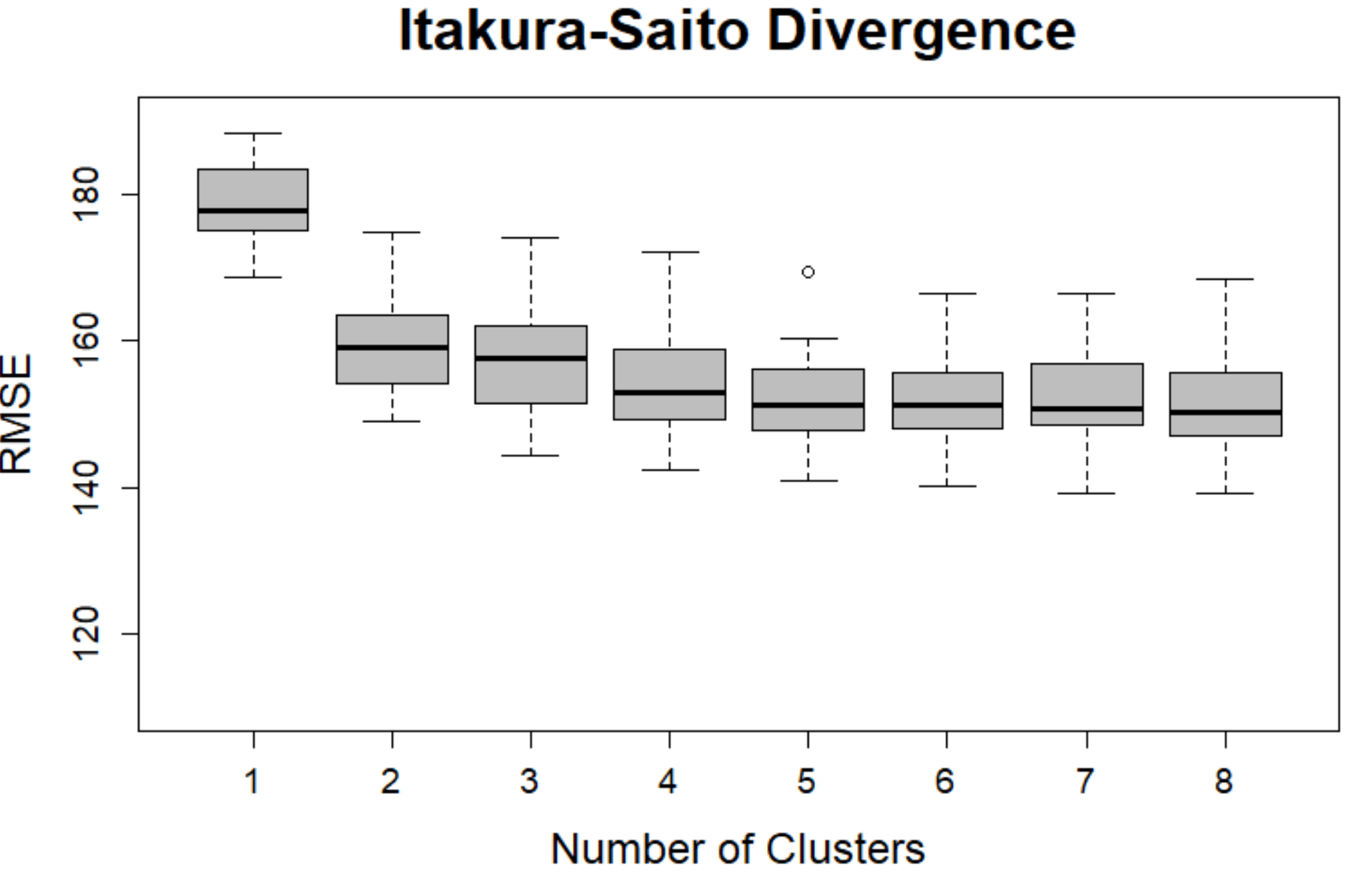}
\includegraphics[width = 6.3cm, height = 4.2cm]{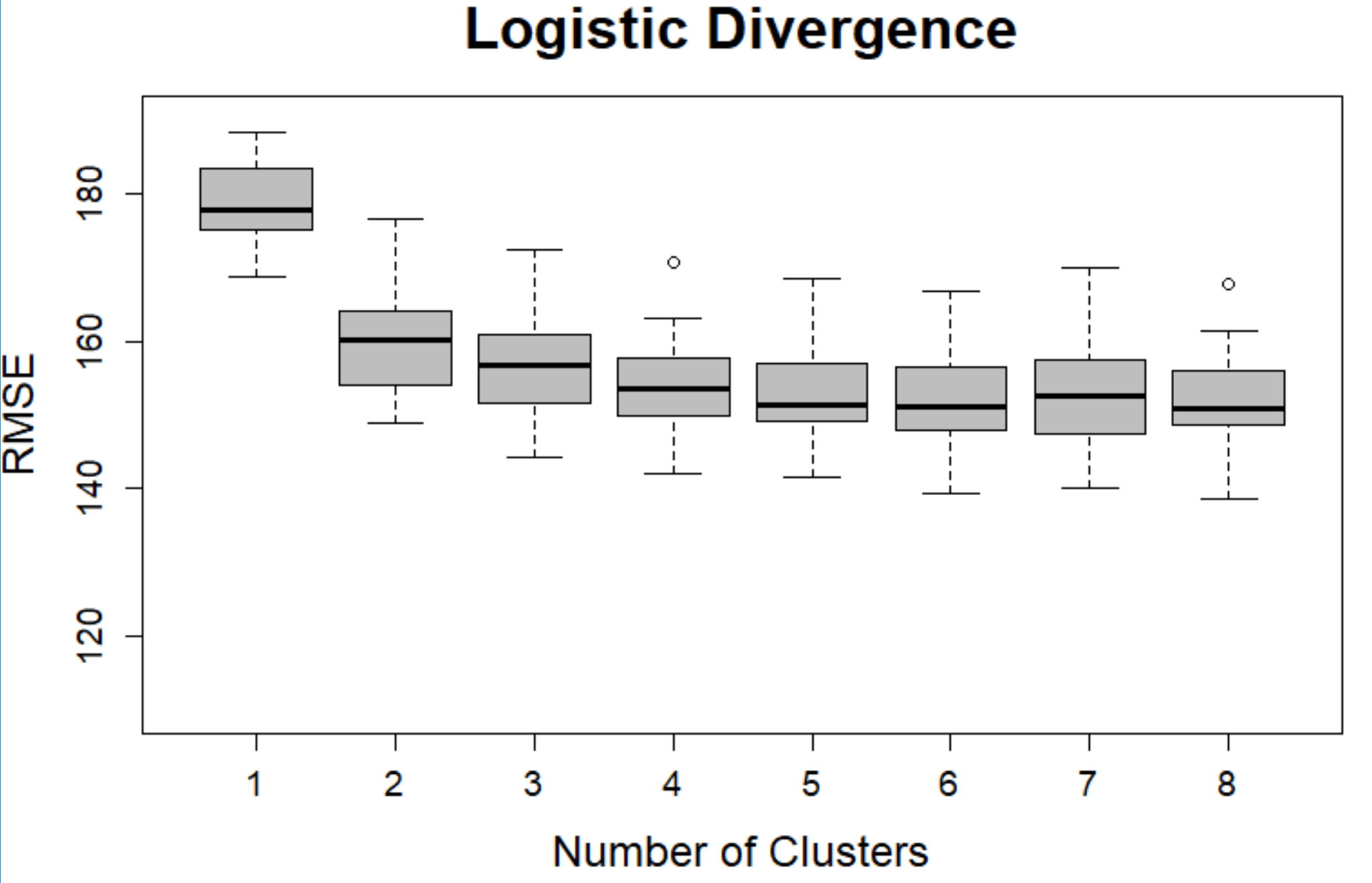}
\includegraphics[width = 6.3cm, height = 4.2cm]{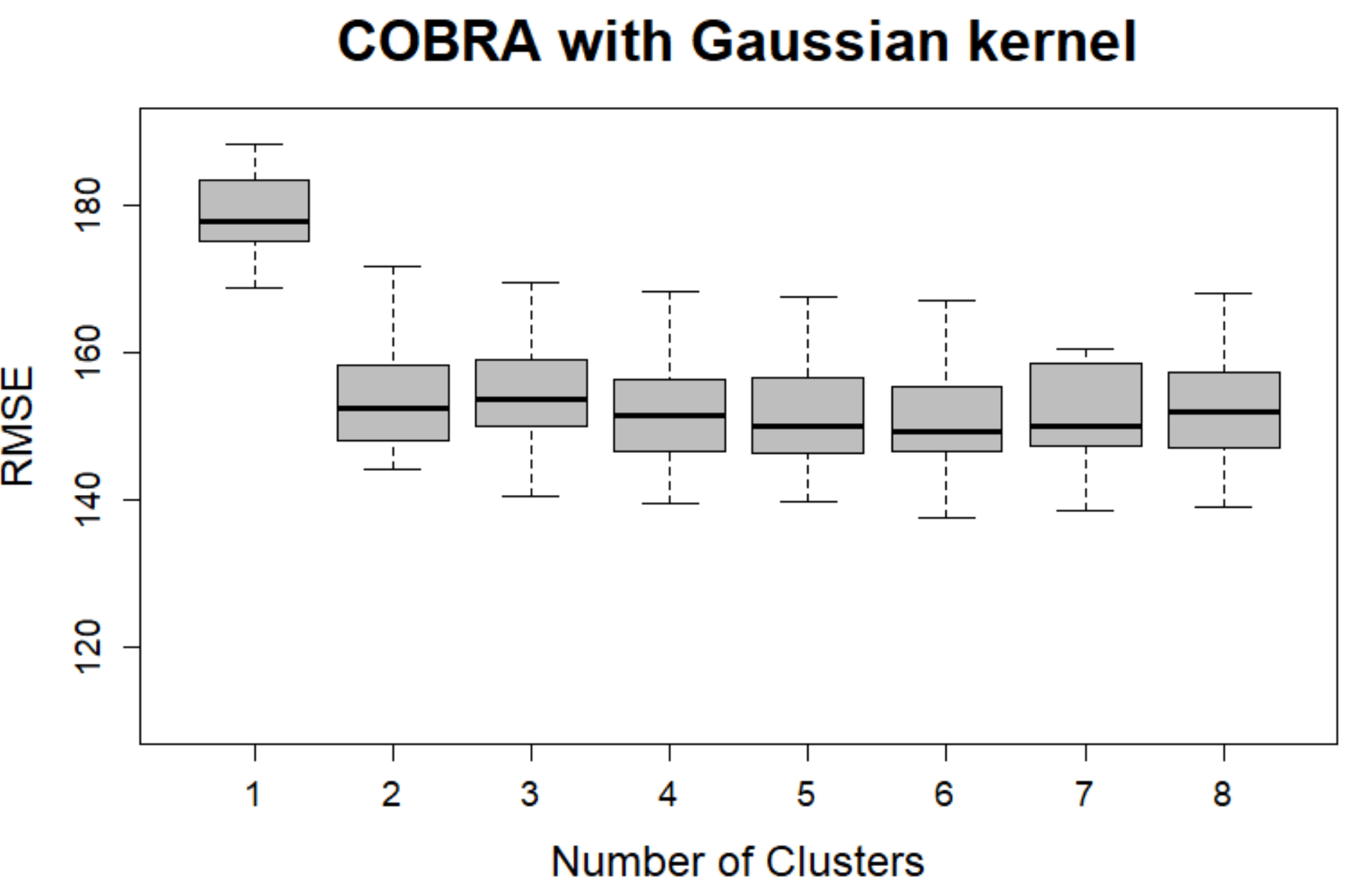}
\includegraphics[width = 6.3cm, height = 4.2cm]{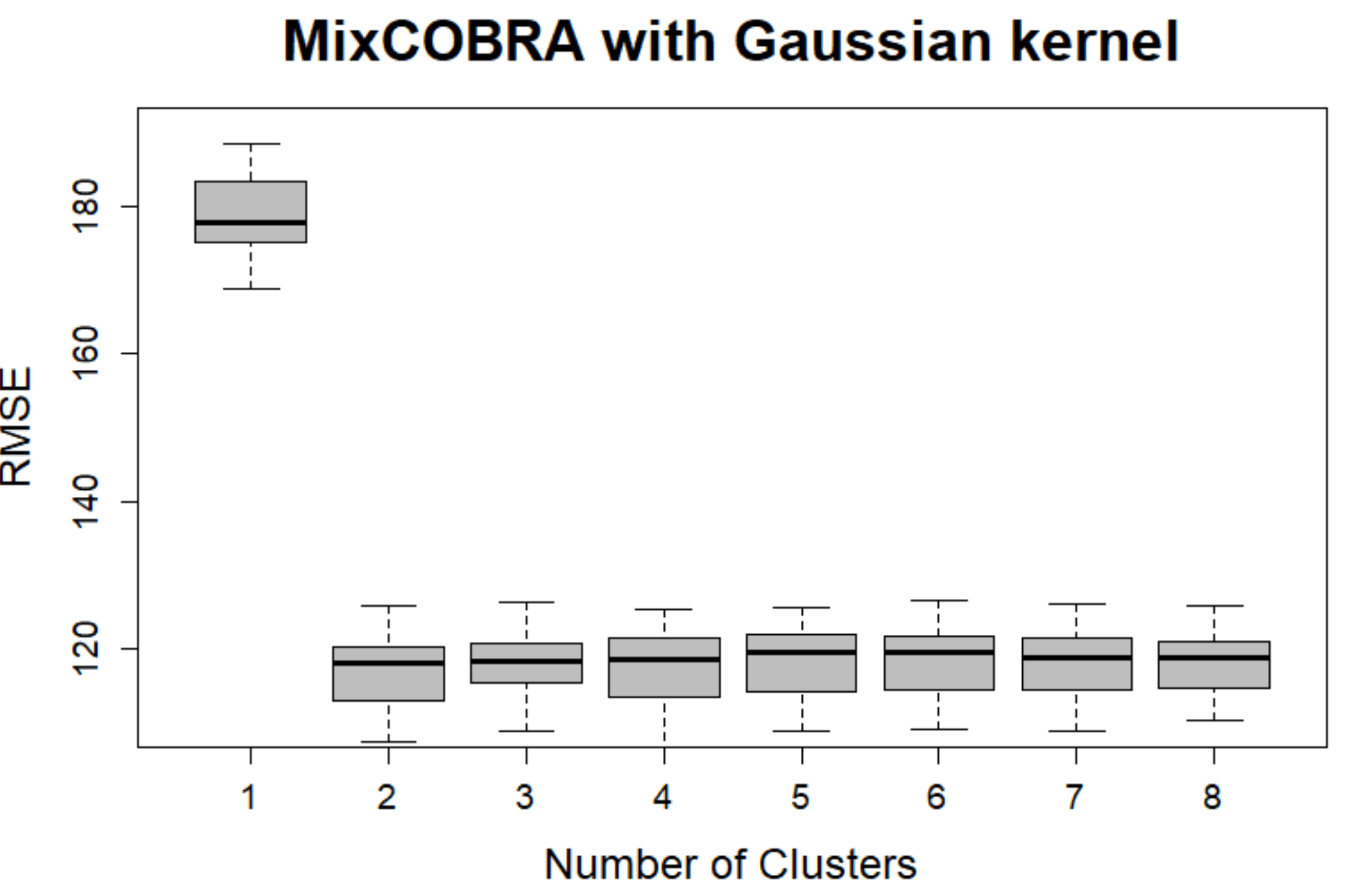}
\caption{Boxplots of RMSE of all the four preliminary models corresponding to the four Bregman divergences in the $K$-step and the resulting models ($Comb_2^R$ and $Comb_3^R$) of the $C$-step, evaluated on Air Compressor data.}
\label{fig:9}
\end{figure}

\section{Conclusion}
\label{sec:conclusion}
The KFC procedure aims to take advantage of the inner groups of input data to provide a consensual aggregation of a set of models fitted in each group built thanks to the K-means algorithm and several Bregman divergences. Simulations using synthetic datasets showed that, in practice, this approach is extremely relevant particularly when groups of unknown distributions belong to the data.
The introduction of several Bregman divergences let automatically captures various shapes of groups.
The KFC procedure brings also relevant improvements for modeling in real-life applications when missing information may induce inner groups. When the number of groups is unknown, which is often the case, cross-validation on the number of groups helps to find the best configurations.

\bibliographystyle{plainnat}

\end{document}